\pdfoutput=1

\documentclass[fleqn,usenatbib]{mnras}

\usepackage{newtxtext,newtxmath}

\usepackage[T1]{fontenc}

\DeclareRobustCommand{\VAN}[3]{#2}
\let\VANthebibliography\thebibliography
\def\thebibliography{\DeclareRobustCommand{\VAN}[3]{##3}\VANthebibliography}

\usepackage{graphicx}	
\usepackage{amsmath}	
\usepackage{caption}
\usepackage[skip=0.5ex]{subcaption}
\usepackage{color,soul}

\title[Early-type galaxies: origin of multiphase gas]{The formation channels of multiphase gas in nearby early-type galaxies}

\author[R. Eskenasy et al.]{
Ryan Eskenasy,$^{1}$\thanks{E-mail: reskenasy@uky.edu (RE)}
Valeria Olivares,$^{1}$
Yuanyuan Su,$^{1}$
and Yuan Li$^{2}$
\\
$^{1}$Department of Physics and Astronomy, University of Kentucky, 505 Rose Street, Lexington, KY 40506, USA\\
$^{2}$Department of Physics, University of North Texas, Denton, TX 76203, USA\\
}

\date{Accepted XXX. Received YYY; in original form ZZZ}

\pubyear{2023}

\begin{document}
\label{firstpage}
\pagerange{\pageref{firstpage}--\pageref{lastpage}}
\maketitle

\begin{abstract}
The processes responsible for the assembly of cold and warm gas in early-type galaxies (ETGs) are not well-understood. We report on the multiwavelength properties of 15 non-central, nearby ($z \leq$ 0.00889) ETGs primarily through Multi-Unit Spectroscopic Explorer (MUSE) and \emph{Chandra} X-ray observations, to address the origin of their multiphase gas. The MUSE data reveals 8/15 sources contain warm ionized gas traced by the \textrm{H}$\alpha$ emission line. The morphology of this gas is found to be filamentary in 3/8 sources: NGC~1266, NGC~4374, and NGC~4684 which is similar to that observed in many group \& cluster-centered galaxies. All \textrm{H}$\alpha$ filamentary sources have X-ray luminosities exceeding the expected emission from the stellar population, suggesting the presence of diffuse hot gas which likely cooled to form the cooler phases.  The morphology of the remaining 5/8 sources are rotating gas disks, not as commonly observed in higher mass systems.  \emph{Chandra} X-ray observations (when available) of the ETGs with rotating \textrm{H}$\alpha$ disks indicate that they are nearly void of hot gas. A mixture of stellar mass loss and external accretion was likely the dominant channel for the cool gas in NGC~4526 and NGC~4710. These ETGs show full kinematic alignment between their stars and gas, and are fast rotators. The \textrm{H}$\alpha$ features within NGC~4191 (clumpy, potentially star-forming ring), NGC~4643 and NGC~5507 (extended structures) along with loosely overlapping stellar and gas populations allow us to attribute external accretion to be the primary formation channel of their cool gas.
\end{abstract}




\section{Introduction}

High energy observations of the intra-cluster medium (ICM) within cool-core clusters reveal the presence of hot ($\sim 10^7$ K), diffuse gas. This gas shines bright in the X-ray, consequently losing energy. As this energy is radiated away, the gas is expected to cool, condense, and ultimately form stars \citep{fabian1994cooling}. However, the measured star-formation rate (SFR) falls significantly short of what steady cooling suggests. 
It is therefore expected that cooling is being balanced by some heating mechanism.

Vast amounts of energy deposited by an active galactic nucleus (AGN) are believed to provide this heating \citep{mcnamara2007heating,olivares2022b}. Such clusters with short cooling times ($<$ 1 Gyr) and low-entropy ($<$ 30 keV cm$^2$) cores show evidence for prominent AGN feedback \citep{cavagnolo2008entropy,pulido2018origin}. Cavities within their X-ray halos, often found to be filled with radio emission \citep{boehringer1993rosat}, indicate that outflows from the AGN disrupt and interact with the surrounding gaseous environment. 

Optical spectra of these clusters with short cooling times and low-entropy cores contain star-formation and extended optical gas nebulae \citep{cavagnolo2008entropy, rafferty2008regulation} depicting the interplay between AGN feedback, star-formation, and cooling. Detailed studies of cool-core clusters \citep{heckman1989dynamical,crawford1999rosat,fabian2003relationship,hatch2006origin,mcdonald2010origin,mcdonald2011star,mcdonald2012optical,mcdonald2012massive,fabian2016hst,hamer2016optical,tremblay2018galaxy,olivares2019ubiquitous} found warm ionized gas nebulae traced by the H$\alpha$ emission line. This warm ($\sim 10^4 $ K) gas is typically found in the form of extended filamentary networks. It is also common for such systems to host cold ($\sim 10-100$ K) molecular gas \citep{edge2001detection, salome2003cold,tremblay2018galaxy,russell2019driving,olivares2019ubiquitous}, often concentrated around the brightest-cluster galaxy (BCG). This cold gas likely serves as fuel for the super massive black hole (SMBH), and can be used as a tool for examining how exactly AGN inject energy into their surrounding environment \citep{rose2019constraining}.

Assessing the origin of the cold and warm gas is a 
logical starting point. 
Simulations of galaxy clusters reveal that thermal instabilities, which cause the hot gas to condense into cooler gas, can be a primary origin \citep{gaspari2012cause, li2014modeling, mccourt2012thermal, sharma2012thermal}. Recent studies have expanded this search to less massive systems, such as galaxy groups. \citet{olivares2022gas} assessed the warm gas reservoirs in galaxy groups within the Complete Local-volume Group Sample (CLoGs) sample \citep{o2017complete}. Their study used high-resolution Multi-Unit Spectroscopic Explorer (MUSE) data to understand the distribution and kinematics of this warm gas. Not only did they find extended H$\alpha$ filaments in the galaxy groups, similar to those observed in clusters, but they also found some of their systems to host rotating compact/extended H$\alpha$ disks within the brightest group galaxy (BGG). For the groups with filamentary warm ionized gas, they attribute thermally unstable cooling of the hot intra-group medium (IGrM) to be the primary origin of the cool gas. 
However, this does not seem to be the only formation channel as some of the galaxy groups with disky H$\alpha$ gas show the gas motion to be coupled to the motion of the stars. This opens the possibility for stellar mass loss to be another origin. Furthermore, many sources show signatures of external interactions and mergers that may have brought in fresh gas.

Multiwavelength observations of galaxy groups and clusters have transformed our understanding of how multiphase gas forms and its connection to the AGN feedback loop. Studies, particularly ones using spatially resolved multiwavelength imaging, attempting to connect this picture to non-central early-type galaxies (ETGs) are sparse. Here, ``non-central" does not inherently imply field galaxies. Rather, ETGs that are neither group or cluster center galaxies. Non-central ETGs are ideal test beds since they have the shallowest gravitational potentials and are thus more dynamically active and susceptible to AGN feedback \citep{wang2019agn}. 
Furthermore, unlike cluster center and group center galaxies, non-central ETGs are not shielded from the ambient ICM and IGrM. The projection of the ICM/IGrM emission in front of or behind the ETGs is minimized, allowing us to probe their interstellar medium (ISM) properties directly.
The origins of multiphase gas are not well understood in these galaxies. \citet{davis2011atlas3d} deduced that kinematic alignment between molecular gas and the stellar component within ATLAS$^{3D}$ ETGs were suggestive of internal origins such as stellar mass loss. Whereas they proposed kinematic misalignment was a signature of external origins, such as galaxy mergers (also see \citet{temi2022probing}). 


In this study, we use archival MUSE data to probe the warm ionized gas within non-central ETGs. \emph{Chandra} X-ray data, when available, is also used to investigate the hot gas occupying their ISM. We refer to the literature to analyze the amount of cold molecular gas within them. In doing so, we attempt to further our understanding of how multiphase gas forms in the lowest mass ETGs. This serves to compliment past studies which did so for high mass systems such as galaxy groups and clusters. Connecting these formation channels, while progressing down the mass hierarchy, acts as a direct probe of AGN feedback and how it can drive galaxy evolution through cosmic time. 

In Section~\ref{sec:sample}, we describe the non-central ETGs within our sample. Section~\ref{sec:data} reviews the data reduction routines for our MUSE and \emph{Chandra} data sets. In Section~\ref{sec:results} we describe the distribution and kinematics of the warm ionized gas as well as the stellar kinematics revealed by our MUSE data. Additionally, we report on the available \emph{Chandra} X-ray data, and past studies describing the cold molecular gas content. Section~\ref{sec:discuss} discusses the connection between the cold molecular gas and warm ionized gas, as well as the various formation channels of the cool gas phases. Our results and interpretation of them are summarized in Section~\ref{sec:conclusion}.

\section{Sample Description}
\label{sec:sample}

Galaxies in our sample are part of the original ATLAS$^{3D}$ survey \citep{cappellari2011atlas3d}. 
This volume-limited catalogue contains a complete sample of 260, morphologically-selected, nearby ($D<42$ Mpc) ETGs and undergoes extensive multiwavelength coverage along with numerical simulations. A primary goal of this project is to characterize the atomic \citep{serra2012atlas3d}, molecular \citep{davis2011atlas3d, young2011atlas3d, davis2013atlas3d} and ionized \citep{davis2011atlas3d} gas content occupying the ISM, and how it is connected with the host galaxy's environment and dynamical structure. The close proximity to these ETGs allows for detailed observations of their stellar and gaseous makeup. 

We select 15 ETGs in the ATLAS$^{3D}$ catalogue
that have been observed with MUSE offering excellent, spatially resolved insight on their optical spectral composition. 8/15 of these non-central ETGs have available \emph{Chandra} X-ray data. The stellar mass of the ETGs in our sample span a broad range of $\sim$ 1$-$39$\times10^{10}$ M$_\odot$. This range falls on the low end of stellar mass estimates of many BGGs ($\sim$ $10^{10}-10^{12}$ M$_\odot$), and is significantly less than typical BCG stellar mass estimates ($\gtrsim10^{11}$ M$_\odot$). 
Table~\ref{tab:atlas} presents the basic properties of our 15 non-central, ETGs.

\begin{table*}

  \caption{General properties of ATLAS$^{3D}$ galaxies within our sample. (1) Name of each non-central ETG in our sample. (2) Redshift taken from the NASA/IPAC Extragalactic Database (NED; \url{http://ned.ipac.caltech.edu/}). (3) Right ascension gathered from NED. (4) Declination gathered from NED. (5) Optical morphology. (6) Logarithm of the half-light radius. Both columns 5 and 6 are gathered from \citet{cappellari2011atlas3d}.}
  \label{tab:atlas}
  \begin{tabular}{cccccc}
    \hline
    Source & z & RA & Dec & Morphology & log R$_e$ \\
    &  & (deg) & (deg) &  & ($^{\prime\prime}$ $/$ kpc)\\
    (1) & (2) & (3) & (4) & (5) & (6)\\
    \hline
    NGC~448 & 0.00636 & 18.818833 & -1.626194 & S0 & 1.05 $/$ 0.12\\ 
    NGC~1266 & 0.00724 & 49.003125 & -2.427361 & S0 & 1.31 $/$ 0.18 \\ 
    NGC~2698 & 0.00634 & 133.902125 & -3.183944 & S0 & 1.10 $/$ 0.17 \\ 
    NGC~4191 & 0.00889 & 183.460019 & 7.2009 & S0 & 1.06 $/$ 0.23 \\ 
    NGC~4264 & 0.00836 & 184.898989 & 5.84675 & S0 & 1.14 $/$ 0.23 \\ 
    NGC~4365 & 0.00415 & 186.117852 & 7.317673 & E & 1.72 $/$ 0.19 \\ 
    NGC~4371 & 0.00311 & 186.23096 & 11.70421 & S0 & 1.47 $/$ 0.13 \\ 
    NGC~4374 & 0.00339 & 186.265597 & 12.886983 & E & 1.72 $/$ 0.17 \\ 
    NGC~4473 & 0.00749 & 187.453628 & 13.42936 & E & 1.43 $/$ 0.26 \\ 
    NGC~4526 & 0.00206 & 188.512856 & 7.699524 & S0 & 1.65 $/$ 0.11 \\ 
    NGC~4643 & 0.00445 & 190.833908 & 1.97827 & S0 & 1.38 $/$ 0.16 \\ 
    NGC~4684 & 0.00520 & 191.822996 & -2.727404 & S0 & 1.32 $/$ 0.18 \\ 
    NGC~4710 & 0.00368 & 192.411795 & 15.165448 & S0 & 1.48 $/$ 0.15 \\ 
    NGC~5507 & 0.00617 & 213.332772 & -3.148885 & S0 & 1.09 $/$ 0.16\\ 
    NGC~5770 & 0.00491 & 223.312567 & 3.959733 & S0 & 1.23 $/$ 0.15 \\ 
    \hline
  \end{tabular}

\end{table*}

\section{Observations \& Data Reduction} 
\label{sec:data}

\subsection{MUSE Analysis} 
\label{subsec:muse}

The MUSE instrument provided optical spectra for our sources. See Table~\ref{tab:muse} for the MUSE observational parameters for each target. The IFU data was reduced via the European Southern Observatory
Recipe Execution Tool (ESOREX v.3.13.1; \citet{weilbacher2020data}). Sky-subtraction was performed with \verb|ZAP| \citep{soto2016zap}, using external sky frames when available. Additionally, we applied the MUSE pipeline's internal sky-subtraction procedure. Object frames were corrected for galactic extinction using the \citet{schlegel1998maps} dust maps.

To fit the stellar continuum and emission lines, a series of code were used. The \verb|AUTOZ| routine \citep{baldry2014galaxy}, using template matching and cross-correlation, was used to estimate the redshift of star light in each IFU pixel. Best-fit velocity dispersions of the stellar component were then calculated using the \verb|VDISPFIT| code. To fit the continuum and emission lines, we adopted the \verb|Platefit| spectral fitting code \citep{brinchmann2004physical, tremonti2004origin}. The fitting procedure used a single Gaussian per emission line. All Balmer lines were tied to have identical velocity and velocity dispersions. Stellar population models and the aforementioned stellar redshift and velocity dispersion products were used to fit the stellar continuum. Optical emission lines were then fitted after subtracting off the underlying continuum, providing H$\alpha$ flux, velocity, and velocity dispersion maps for each source.

To clean the H$\alpha$ maps of background noise and contamination, we excluded pixels with S/N $<$ 7, and with gas velocity dispersions $<$ 40 km s$^{-1}$. Additionally, we remove pixels that have 1$\sigma$ velocity errors greater than a given threshold. These thresholds were determined on a source-by-source basis, by calculating the average velocity dispersion error in an offset, background region of the corresponding maps.

\begin{table*}

  \caption{MUSE observations of non-central ETGs in our sample. (1) Target name. (2) ESO program identification. (3) Exposure time along with number of frames. (4) Average FWHM seeing calculated over full exposure time (averaged over all frames). (5) H$\alpha$ emission line observed in MUSE data cube's spectrum (check if observed, dash if absent).}
  \label{tab:muse}
  \begin{tabular}{ccccc}
    \hline
    Source & Program ID & Exposure time & Average seeing & H$\alpha$ emission line \\
     &  & (sec) & ($^{\prime\prime}$ $/$ kpc) & \\
     (1) & (2) & (3) & (4) & (5) \\
    \hline
    NGC~448 & 094.B-0225(A) & 1291 (8) &  0.75 $/$ 0.08 & -\\
    NGC~1266 & 0102.B-0617(A) & 600 (4) & 0.77 $/$ 0.11 & $\checkmark$\\
    NGC~2698  & 096.B-0449(A) & 678 (12) &  0.79 $/$ 0.12 & - \\
    NGC~4191 & 095.B-0686(A) & 840 (3) & 0.91 $/$ 0.19 & $\checkmark$\\
    NGC~4264 & 094.B-0241(A) & 780 (8) &  1.3 $/$ 0.26 & - \\
    NGC~4365 & 094.B-0225(A) & 1251 (2) &  1.4 $/$ 0.16 & - \\
    NGC~4371 & 60.A-9313(A) & 600 (7) &  0.82 $/$ 0.07 & - \\
    NGC~4374 & 0102.B-0048(A) & 600 (4) & 0.89 $/$ 0.09 & $\checkmark$\\
    NGC~4473 & 095.B-0686(A) & 810 (3) &  1.2 $/$ 0.22 & - \\
    NGC~4526 & 097.D-0408(A) & 590 (4) & 0.57 $/$ 0.04 & $\checkmark$\\
    NGC~4643 & 097.B-0640(A) & 480 (8) & 0.70 $/$ 0.08 & $\checkmark$\\
    NGC~4684 & 096.B-0449(A) & 678 (14), 121 (1)	 & 0.85 $/$ 0.11 & $\checkmark$\\
    NGC~4710 & 60.A-9307(A) & 600 (4) & 0.64 $/$ 0.06 & $\checkmark$\\
    NGC~5507 & 096.B-0449(A) & 678 (12) & 1.1 $/$ 0.17 & $\checkmark$\\
    NGC~5770 & 096.B-0449(A) & 678 (12) &  1.2 $/$ 0.14 & - \\
    \hline
  \end{tabular}

\end{table*}

\subsubsection{Stellar kinematics}
\label{subsec:stellar}

Kinematics of the stellar component were measured using the Galaxy IFU Spectroscopy Tool (GIST) pipeline \citep{2019A&A...628A.117B}. The science-ready MUSE IFU data was spatially binned following the Voronoi tesselation routine in \citet{cappellari2003adaptive}. Spectra were then logarithmically re-binned, to which they were again spatially binned. To extract the stellar kinematics, the spectra were fed to the \verb|pPXF| routine \citep{cappellari2004parametric}. This routine produces line-of-sight velocity and velocity dispersion maps for a given rest-frame wavelength range and initial velocity dispersion guess. Foreground stars were manually removed after feeding the MUSE IFU data through the GIST pipeline.

\subsection{Chandra Analysis
\label{subsec:chandra}}


A subset of our sample was observed with the \emph{Chandra} X-ray observatory (see Table~\ref{tab:chandra} for the observational log). The data was reduced using CIAO version 4.14 \citep{fruscione2006ciao} and CALDB version 4.10.2.

Level 1 data was first reprocessed using \verb|chandra_repro|. Out-of-time events from X-ray bright sources were then corrected for with \verb|acisreadcorr|. The resulting data set was cleared of anomalous ACIS background flares using the \verb|deflare| routine. Individual observations were reprojected to a matching coordinate with \verb|reproject_obs| where the following event files were then merged. Background files were created via the \verb|blanksky| procedure, for all sources except for NGC~4374\footnote{NGC~4374 resides near the center of the Virgo cluster. To minimize the impact of the Virgo cluster emission, we use a local background from an annulus of 3 R$_e$$-$ 4 R$_e$ for estimating its background.}.
Exposure maps were generated using the \verb|flux_obs| routine. \verb|wavdetect| was then used to identify and remove point sources. Areas that were originally occupied by point source emission were replaced with a region adjacent to the point source so that it contains the ambient ISM emission and all sources of background. The resulting 0.5-2.0 keV image was created by subtracting the count map by the background file, followed by a normalization with the exposure map.

Background subtracted net count rates were collected from a circular region within a radius of 1 R$_e$. Using the PIMMS Mission Count Rate Simulator\footnote{\url{https://cxc.harvard.edu/toolkit/pimms.jsp}}, we converted the net count rate to an energy flux. We assumed the Plasma/APEC model with a metallicity of 1 $Z_{\odot}$ and $kT=0.5$ keV\footnote{We assume $kT=0.87$ keV and $kT=0.76$ keV for NGC~1266 and NGC~4374, respectively, as reported in \citealt{su2015scatter}.} (for typical ETGs hot ISM, \citealt{su2013investigating}). To generalize the impact on the X-ray luminosity by choosing different $kT$ and metallicities, we quantified the total spread in NGC~5507's resulting $L_X$ assuming (1) $kT=0.5$ keV and $Z=1$ $Z_{\odot}$ (2) $kT=0.5$ keV and $Z=0.5$ $Z_{\odot}$ (3) $kT=0.3$ keV and $Z=1$ $Z_{\odot}$. We chose NGC~5507 as a reference given its intermediate stellar mass and Chandra cycle compared with other ETGs in our sample. The total percent uncertainty covering these parameter-varying estimates is $\sim20\%$.
HI column densities were gathered from the HEASARC \href{https://heasarc.gsfc.nasa.gov/cgi-bin/Tools/w3nh/w3nh.pl}{$N_H$ tool} \citep{bekhti2016hi4pi}.
We also considered the scenario that most of their diffuse X-ray emission originate from unresolved X-ray binaries, for which we assume a power-law model with $\textrm{photon index}=1.7$. The resulting $L_{\rm X}$ estimates by assuming APEC or power-law models are shown in Fig.~\ref{fig:Lx}. Their discrepancies are within 20\%.

NGC~4684 was serendipitously detected by a Chandra ACIS-I observation where it falls onto the far edge of the detector (8.5$^{\prime}$ from the pointing position). At this location the effective area has declined sharply and the point spread function is distorted. PIMMS conversions from count rates to energy fluxes hold best for on-axis sources. Thus, instead of estimating its X-ray flux based on a count rate, we produced an image in units of photon s$^{-1}$ cm$^{-2}$ using \verb|flux_obs| which takes into account the difference in Chandra cycle and off-axis position. Then we used \verb|Xspec| version 12.12.1 to convert photon s$^{-1}$ cm$^{-2}$ into erg s$^{-1}$ cm$^{-2}$, for which we assume an absorbed thermal APEC model with a metallicity of 1 $Z_{\odot}$ and $kT=0.5$ keV or a power-law model. 


 

\begin{table}

  \caption{\emph{Chandra} observations. (1) Target name. (2) Instrument used during observation. (3) Observation ID. (4) Net exposure time integrated over all observations.}
  \label{tab:chandra}
  \begin{tabular}{cccc}
    \hline
    Source & Instrument & OBSID & Net exposure time  \\
     &  &  & (ksec)\\
    (1) & (2) & (3) & (4)\\
    \hline
    NGC~1266 & ACIS-S &11578, 19498, 19896 & 148.1 \\
    & & 11578, 19498, 19896&\\
    NGC~4365 & ACIS-S &2015, 5921, 5922 & 195.8 \\
    & & 5923, 5924, 7224&\\
    NGC~4374 & ACIS-S	 & 401, 803, 5908 & 884.1 \\
    & & 6131, 20539, 20540&\\
    & & 20541, 20542, 20543&\\
    & & 21845, 21852, 21867&\\
    & & 22113, 22126, 22127&\\
    & & 22128, 22142, 22143&\\
    & & 22144, 22153, 22163&\\
    & & 22164, 22166, 22174&\\
    & & 22175, 22176, 22177&\\
    & & 22195, 22196&\\
    NGC~4473 & ACIS-S & 4688, 11736, 12209 & 107.6 \\
    NGC~4526 & ACIS-S & 3925 & 43.5 \\
    NGC~4684 & ACIS-I & 15190 & 28.8 \\
    NGC~4710 & ACIS-S	 & 9512 & 29.8 \\
    NGC~5507 & ACIS-S & 358, 26398 & 18.7 \\
    \hline
  \end{tabular}

\end{table}

\section{Results} 
\label{sec:results}

\subsection{Warm ionized gas morphology and kinematics} 
\label{sec:warm}

The H$\alpha$ emission line, which traces the warm ionized gas, is observed in the MUSE spectrum for 8/15 sources in our sample. Flux, average velocity, and velocity dispersion maps of this H$\alpha$ gas are presented in
Fig.~\ref{fig:Ha_Stell}. The warm ionized gas morphology, luminosity, and kinematic position angle for these sources are summarized in Table~\ref{tab:data}.

\begin{figure*}
  \centering 
  \includegraphics[width=0.7\linewidth]{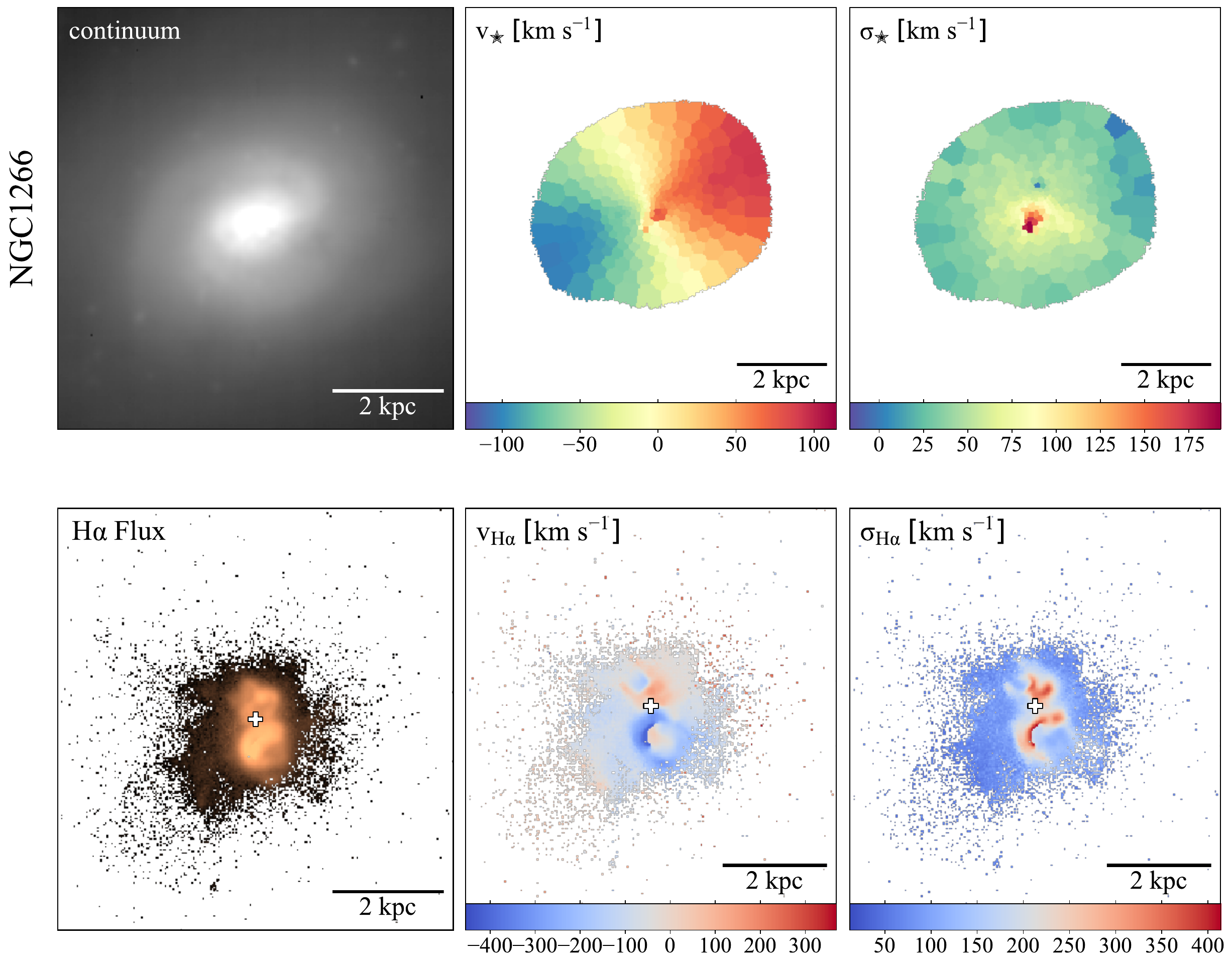}%
  \caption{Stellar and warm ionized gas maps extracted from \textbf{NGC~1266's} MUSE data cube. \textbf{Top row}: Log-scale MUSE optical continuum image (left), stellar velocity (center), and stellar velocity dispersion (right). \textbf{Bottom row}: Log-scale H$\alpha$ flux map (left), H$\alpha$ velocity (center), and H$\alpha$ velocity dispersion (right). White cross marks position of the optical center.}
  \label{fig:Ha_Stell}
\end{figure*}

\begin{figure*}
  \ContinuedFloat 
  \centering 
  \includegraphics[width=0.7\linewidth]{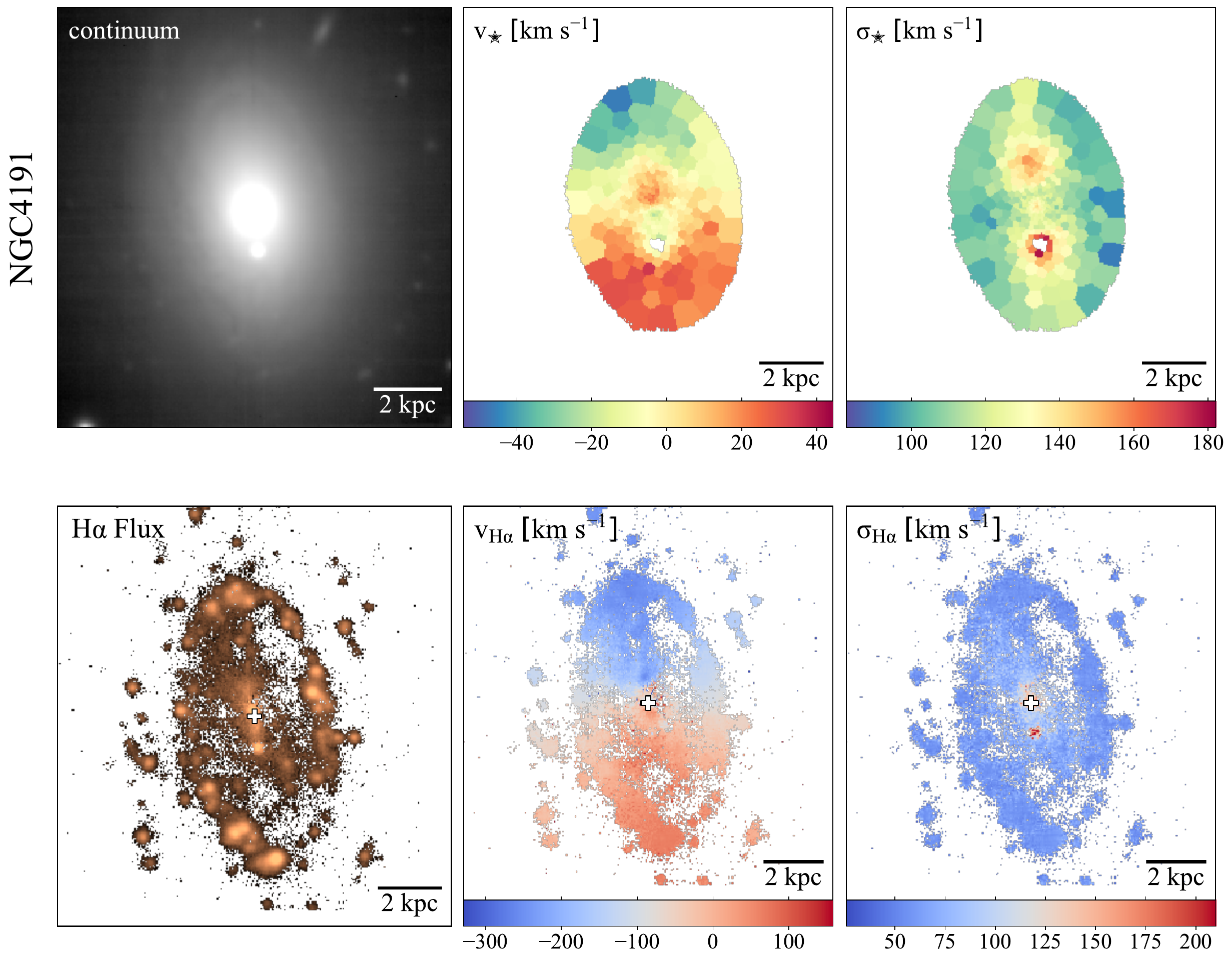} 
  \caption[cont]{\textbf{(continued) NGC~4191}}
\end{figure*} 

\begin{figure*}
  \ContinuedFloat 
  \centering 
  \includegraphics[width=0.7\linewidth]{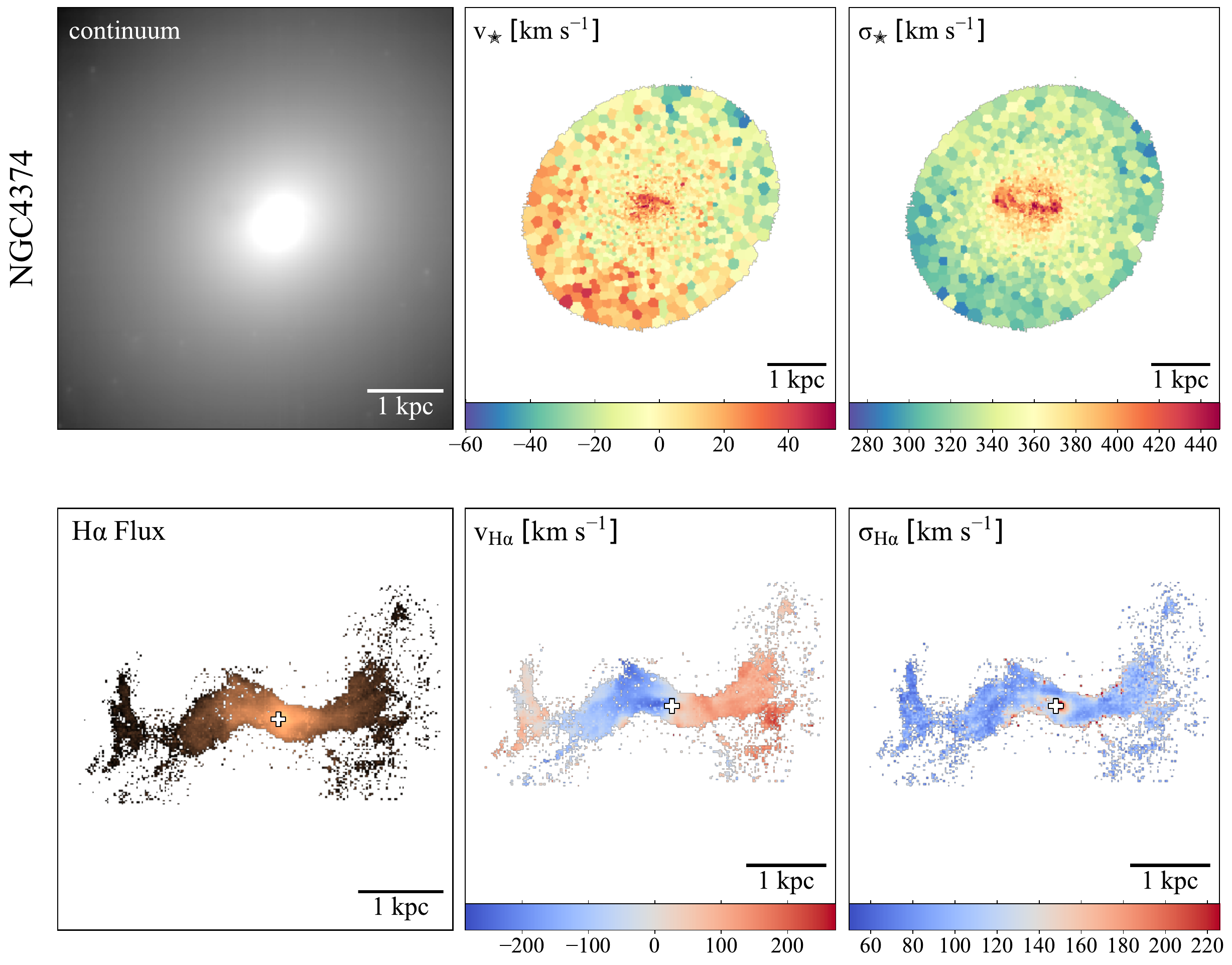} 
  \caption[]{\textbf{(continued) NGC~4374}}
\end{figure*} 

\begin{figure*}
  \ContinuedFloat 
  \centering 
  \includegraphics[width=0.7\linewidth]{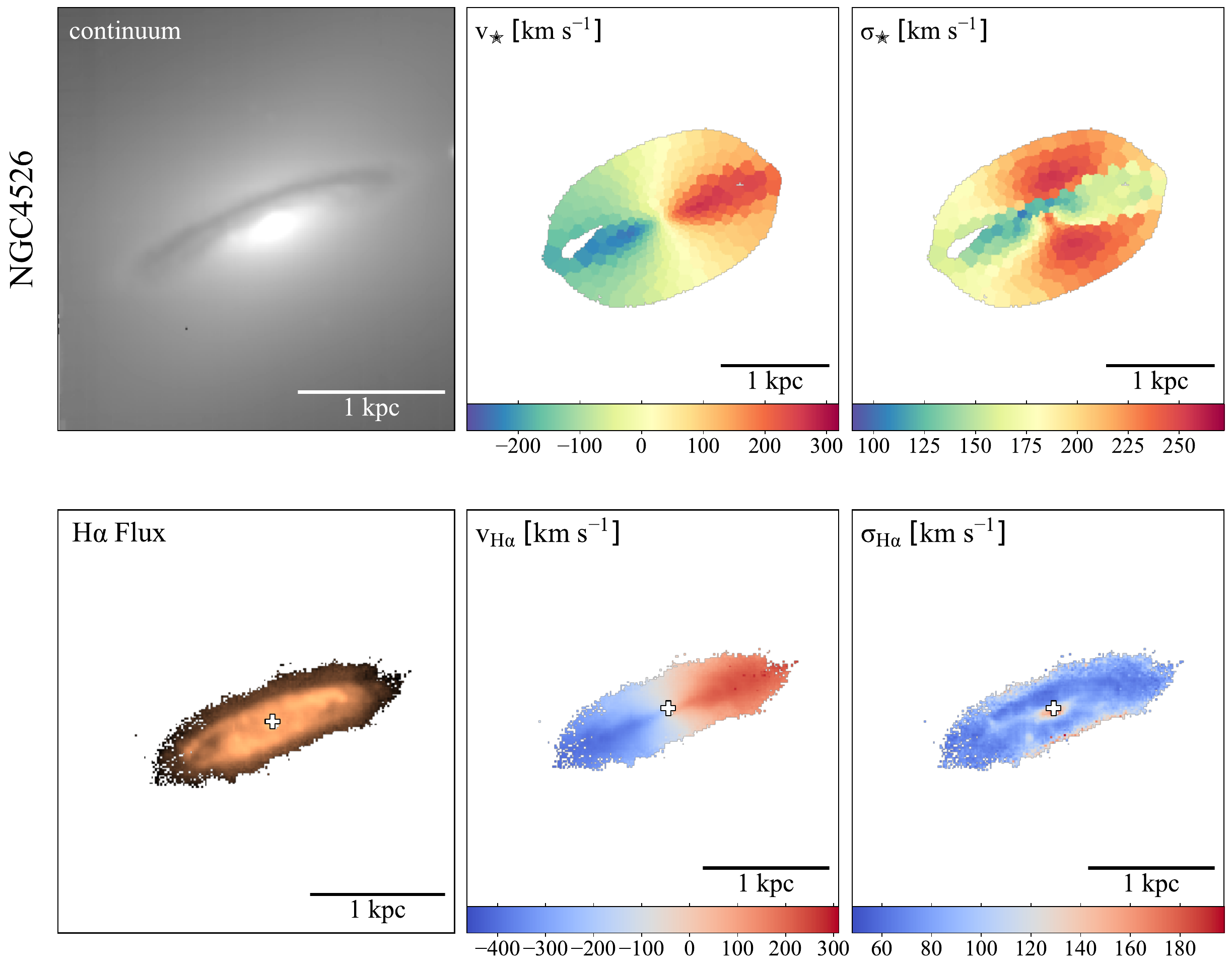} 
  \caption[]{\textbf{(continued) NGC~4526}}
\end{figure*} 

\begin{figure*}
  \ContinuedFloat 
  \centering 
  \includegraphics[width=0.7\linewidth]{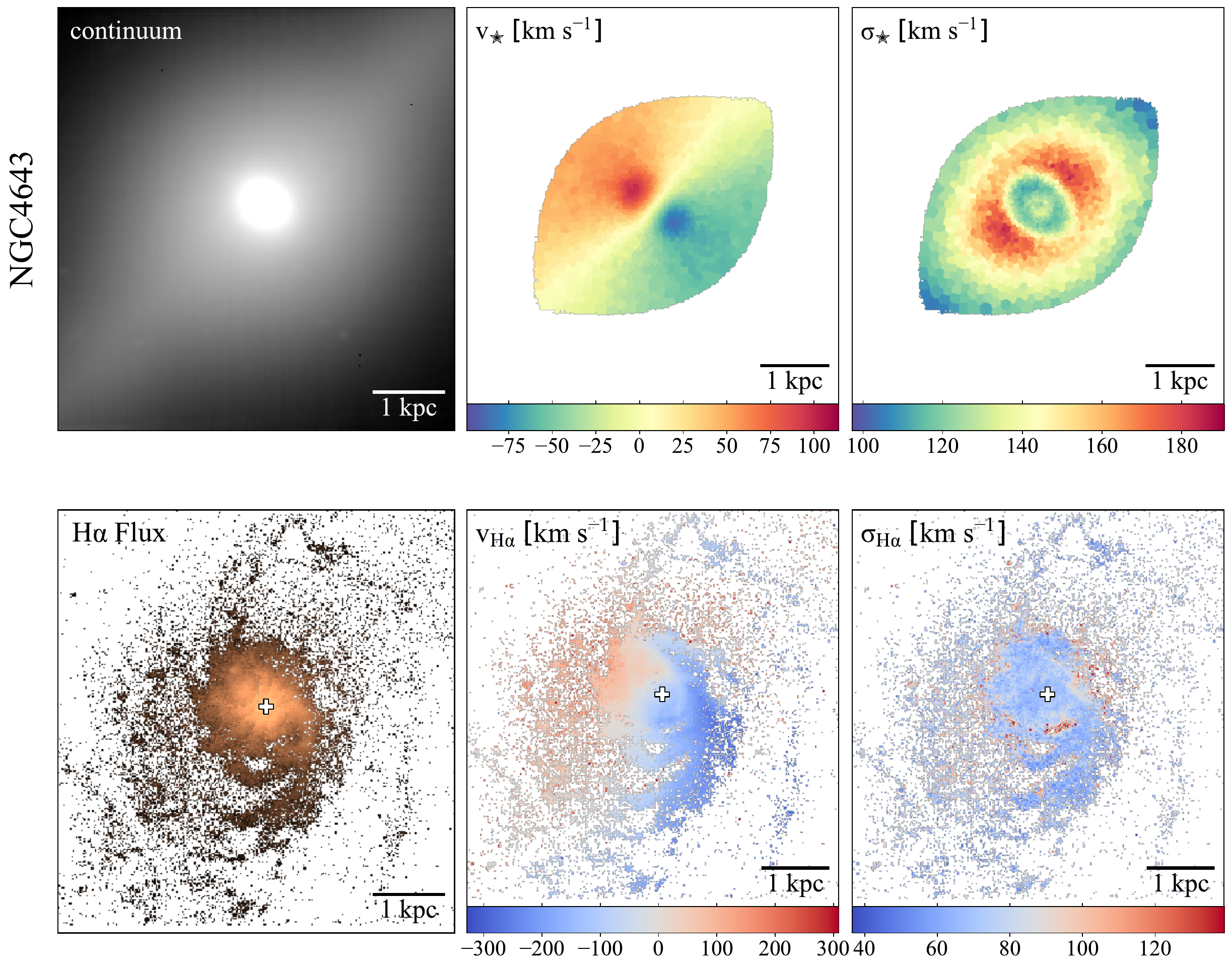} 
  \caption[]{\textbf{(continued) NGC~4643}.}
\end{figure*} 

\begin{figure*}
  \ContinuedFloat 
  \centering 
  \includegraphics[width=0.7\linewidth]{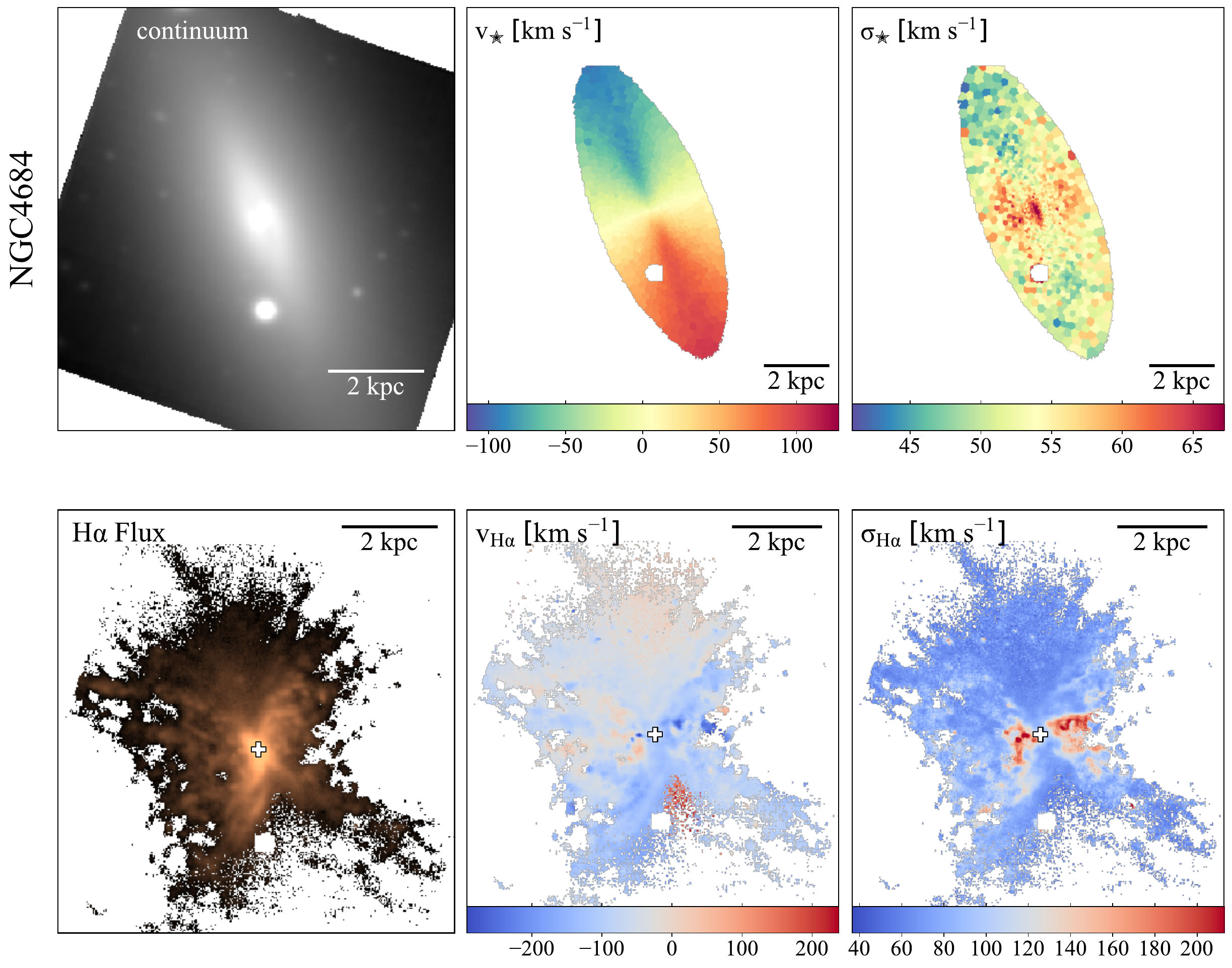} 
  \caption[]{\textbf{(continued) NGC~4684}}
\end{figure*} 

\begin{figure*}
  \ContinuedFloat 
  \centering 
  \includegraphics[width=0.7\linewidth]{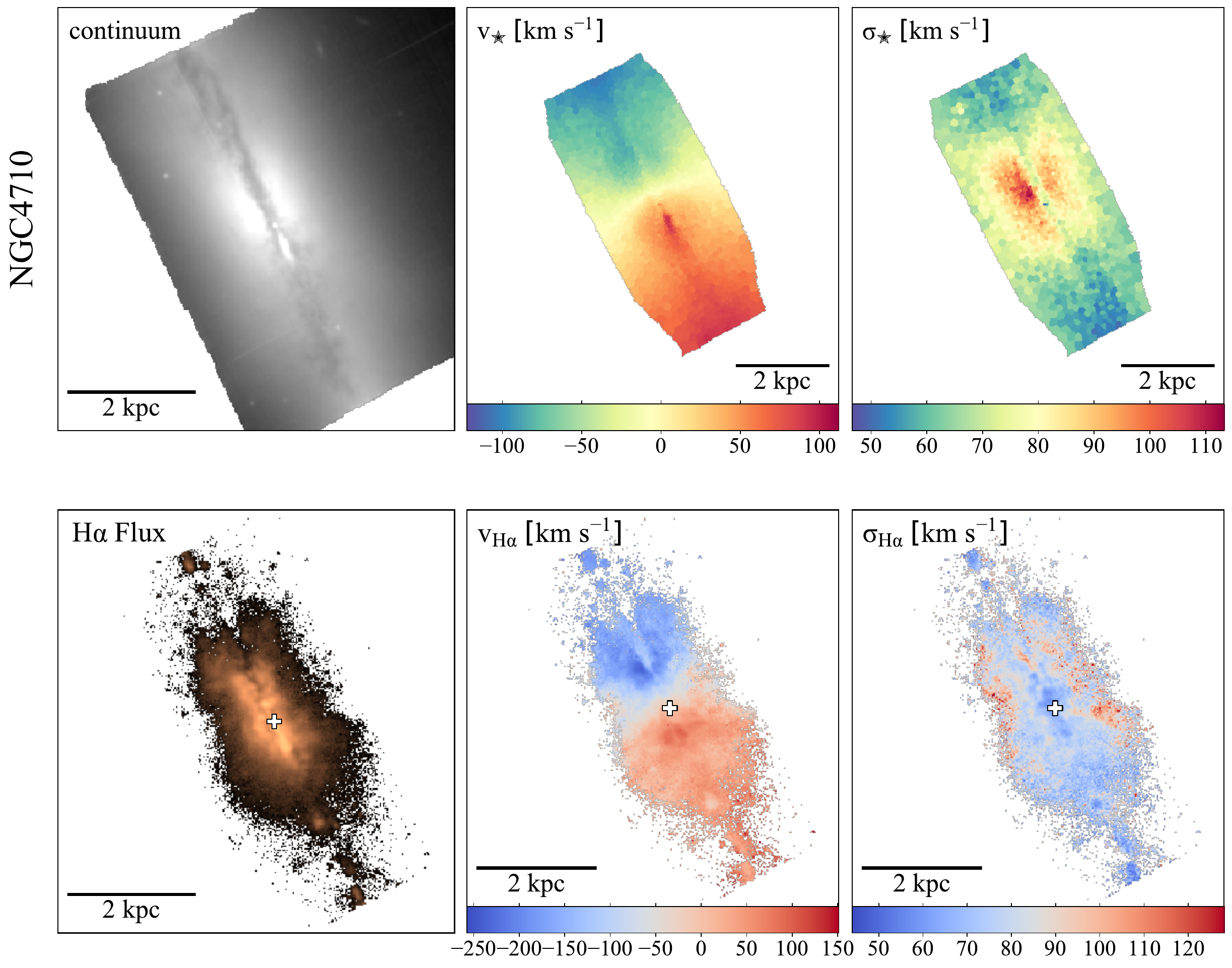} 
  \caption[]{\textbf{(continued) NGC~4710}}
\end{figure*} 

\begin{figure*}
  \ContinuedFloat 
  \centering 
  \includegraphics[width=0.7\linewidth]{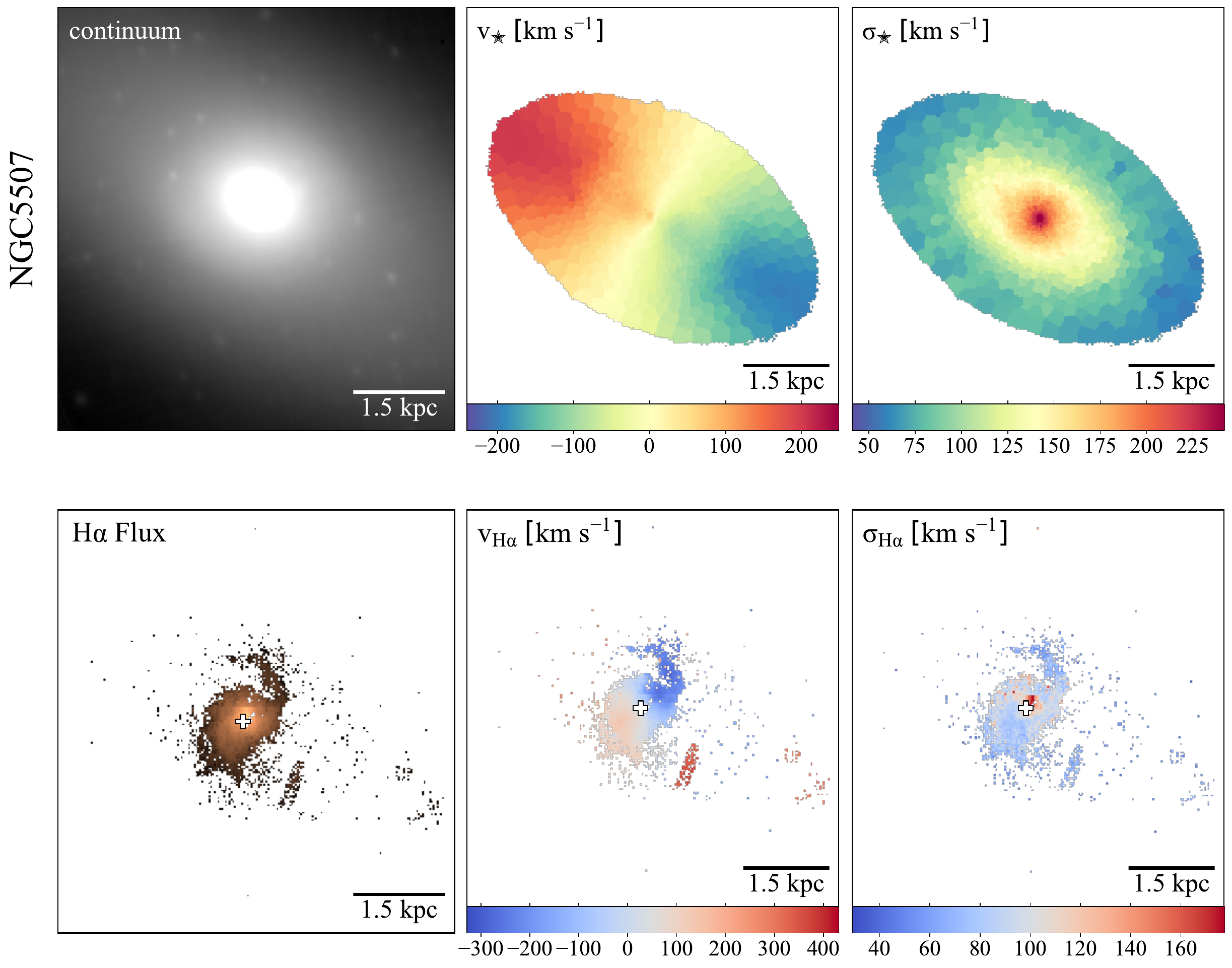} 
  \caption[]{\textbf{(continued) NGC~5507}}
\end{figure*} 

Each of these eight, H$\alpha$-bright sources are classified as either ``rotating disks" or ``filaments" based on the morphology and kinematics of their H$\alpha$ gas (as observed in Fig.~\ref{fig:Ha_Stell}). Rotating disks are classified as sources with H$\alpha$ velocity fields that display ordered rotation around the optical center of the galaxy, as well as lack filaments. Filaments are classified as sources with irregular H$\alpha$ branches and sub-structures. The H$\alpha$ classification for each source is listed in Table~\ref{tab:data}.

Five sources (NGC~4191, NGC~4526, NGC~4643, NGC~4710, and NGC~5507) are classified as rotating H$\alpha$ disks. The sizes of these rotating disks, defined as the length of the projected semi-major axis, span $\sim1-5$ kpc. NGC~4191's rotating H$\alpha$ disk is the largest ($\sim5$ kpc) out of the five, and is comprised of smaller-scale H$\alpha$ clumps that form a ring-like structure around the center. NGC~4526 has a highly uniform H$\alpha$ disk which extends to $\sim1$ kpc. NGC~4643's disk ($\sim1$ kpc) has prominent spirals tracing the rotation of the H$\alpha$ gas. NGC~4710 ($\sim2$ kpc) is a target viewed edge-on with a rotating H$\alpha$ disk that is seemingly volume-filling. NGC~5507's rotating disk has a size of $\sim1$ kpc and potentially hosts an unresolved clump in the southwest direction.

The three other ETGs (NGC~1266, NGC~4374, and NGC~4684) that contain the H$\alpha$ emission line in their optical spectra are classified as filamentary H$\alpha$ sources. The sizes of these filaments, defined as the projected distance from the center of the galaxy to the edge of the most extended filament, range from $\sim 2-5$ kpc. NGC~1266 displays more compact, knotty clumps ($\sim 2$ kpc radially) along with a prominent whorl south of the SMBH. These components comprise shell-like gas structures around its center (see Sec.~\ref{sec:outflows} 
for a detailed description of the possible physical drivers for NGC~1266's distinct warm ionized gas structure).
NGC~4374 has extended filaments ($\sim2$ kpc radially) east and west of its SMBH. Both filaments form a curve close to the center and extend into wider ``wings" at radii $\gtrsim1$ kpc. NGC~4684's extended H$\alpha$ filaments ($\sim5$ kpc radially) are the largest of these ETGs, and form a unique ``X" shape. We note that while NGC~4374 is classified as a filamentary source, it also shows rotation in its H$\alpha$ velocity fields.

\subsection{Stellar kinematics} 
\label{sec:stell_kin}

For the early-type galaxies in our sample that host H$\alpha$ gas, we measured the kinematics of their stellar populations (see Section~\ref{subsec:stellar}). Stellar velocity and velocity dispersion maps are presented in Fig.~\ref{fig:Ha_Stell}.

Six of our H$\alpha$ emitting galaxies (NGC~1266, NGC~4526, NGC~4643, NGC~4684, NGC~4710, NGC~5507) exhibit stellar motion dominated by ordered rotation, as determined through visual inspection of their velocity fields. NGC~4191 shows minimal ordered rotation while NGC~4374's stellar velocity field is nearly absent of rotation. This is consistent with the classifications made in \citet{krajnovic2011atlas3d} where stellar kinematics of the ATLAS$^{3D}$ galaxies were extracted from the SAURON integral-field spectrograph \citep{bacon2001sauron}. In this study, they similarly deduced stellar motion dominated by ordered rotation, via kinemetry, in the six galaxies mentioned above (and conversely for NGC~4191 and NGC~4374).

\subsection{X-ray analysis}
\label{chandra_results}

\begin{figure*}
    \centering
    \includegraphics[width=\linewidth]{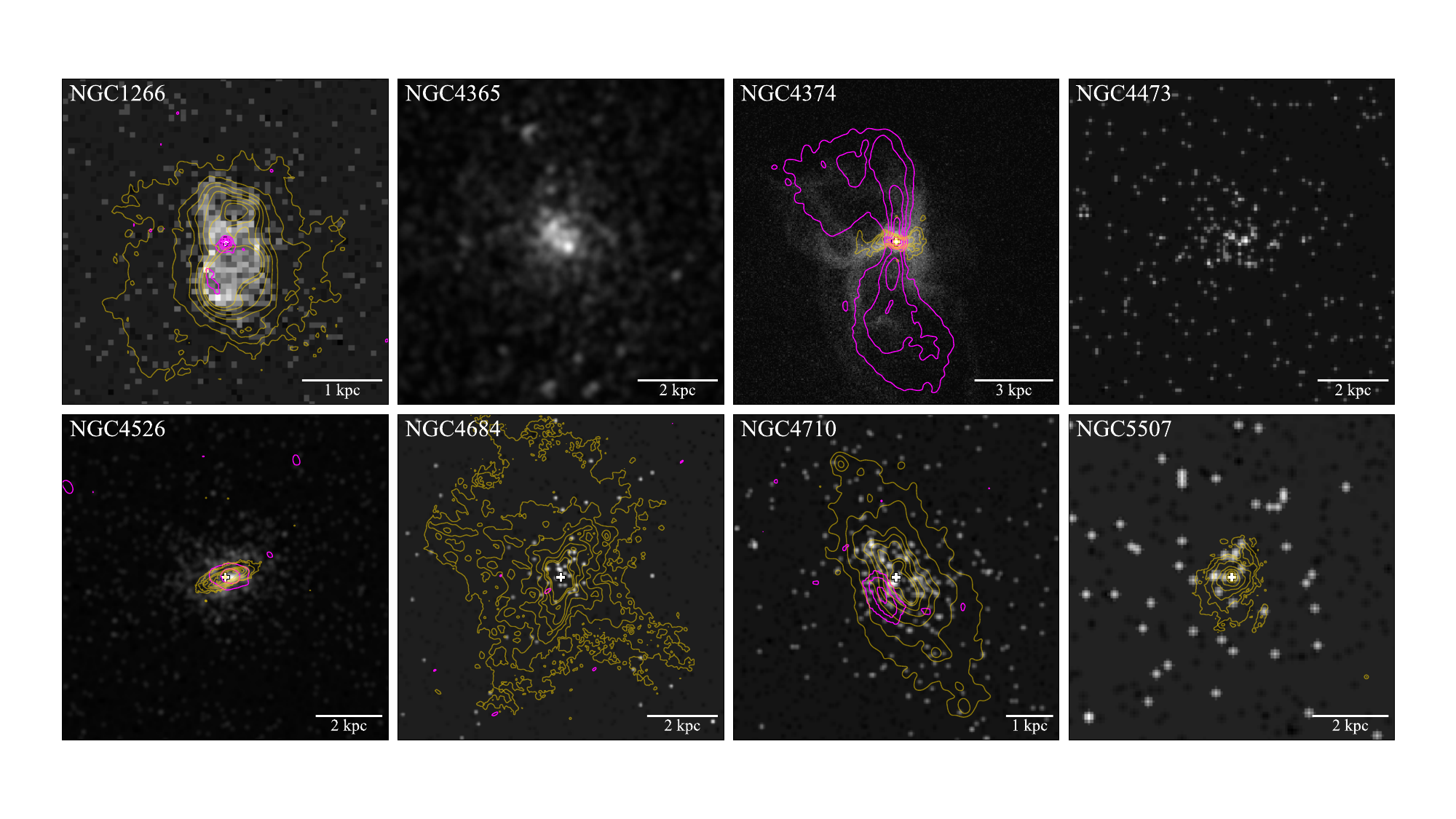}
    \caption{0.5-2.0 keV (log scale) images are shown for all sources in our sample that were observed with \emph{Chandra}. Over plotted in yellow and pink contours are the H$\alpha$ and 4.89 GHz radio emission, respectively (at levels of 3$\sigma$, 10$\sigma$, 25$\sigma$, 50$\sigma$, 100$\sigma$, 200$\sigma$, and 300$\sigma$ where $\sigma$ is the rms noise level). The name of the galaxy is given in the top left of each panel. The physical scale is shown on the bottom right of each panel.}
    \label{fig:xray_radio}
\end{figure*}

Fig.~\ref{fig:xray_radio} displays the 0.5-2.0 keV images for all sources in our sample that were observed with \emph{Chandra}. Overlaid are pink contours depicting the 4.89 GHz radio emission captured with the VLA\footnote{Image credit: NRAO/VLA Archive Survey: \url{http://www.aoc.nrao.edu/~vlbacald/ArchIndex.shtml}}. For those  with H$\alpha$ emission detected in their MUSE data, we overlay yellow contours depicting their H$\alpha$ flux. The 0.5-2.0 keV luminosities (measured within 1 R$_e$), for our sources with H$\alpha$ emission and available \emph{Chandra} data range from $\sim$ 0.4$-$46$\times 10^{39}$ erg s$^{-1}$, and are listed in Table~\ref{tab:data}. We calculate X-ray luminosities for each source as described in Sec. \ref{subsec:chandra}

NGC~1266 (148.1 ksec) and NGC~4374 (884.1 ksec) had the deepest exposure times with \emph{Chandra}. NGC~1266's 0.5-2.0 keV X-ray emission displays sharp discontinuities along its perimeter. An X-ray cavity is potentially shown north of the nucleus. The 4.89 GHz radio emission is found in the nuclear region, along with an extended structure along the southeast direction. While not depicted in the archival VLA data, \citet{nyland2016atlas3d} found resolved radio lobes/jets using more recent 5 GHz VLA observations. NGC~4374 displays clear X-ray cavities that are filled with radio emission. NGC~4684 displays weak 4.89 GHz radio emission, while NGC~4365, NGC~4473, and NGC~5507 are nearly absent of radio emission.

\begin{table*}

  \caption{Multiwavelength results of the H$\alpha$-emitting ETGs in our sample. (1) Name of H$\alpha$-emitting non-central ETG in our sample. (2) Morphological classification of the H$\alpha$ nebulae. (3) Total H$\alpha$ luminosity. (4) Kinematic position angle of the H$\alpha$ velocity field. (5) Logarithm of the molecular gas mass. (6) Stellar mass. (7) Kinematic position angle of the stellar velocity field. (8) Difference between the H$\alpha$ and stellar kinematic position angles. (9) 0.5-2.0 keV luminosity measured within 1 R$_e$ (assuming APEC model).}
  \label{tab:data}
  \begin{tabular}{ccccccccc}
    \hline
    Source & H$\alpha$ morphology & $L_{\mathrm{H\alpha}}$ & PA$_{\mathrm{H\alpha}}$   & log M$(\mathrm{H_2})$ & M$_{\star}$ & PA$_{\star}$  & $\Delta \textrm{PA}$ & $L_X$ \\
     &  & ($10^{40} \textrm{ erg } \textrm{s}^{-1}$) & (deg) & (M$_{\odot}$) & ($10^{10}$ M$_{\odot}$) & (deg) & (deg) & ($10^{39} \textrm{ erg } \textrm{s}^{-1}$)\\
     (1) & (2) & (3) & (4) & (5) & (6) & (7) & (8) & (9)\\
    \hline
    NGC~1266 & filaments & 1.2 $\pm$ 0.1 & 165.6 $\pm$ 0.6 & 9.28 $\pm$ 0.01 & 2.58 & 114.9 $\pm$ 0.9 & 50.7 & 3.51 $\pm$ 0.11  \\
    NGC~4191 & rotating disk & 0.60 $\pm$ 0.02 & 13.8 $\pm$ 0.5 & $<7.94$ & 5.06 & 11.8 $\pm$ 6.3 & 2.1 & -  \\
    NGC~4374 & filaments & 0.439 $\pm$ 0.003 & 81.3 $\pm$ 0.5 & $<7.23$ & 38.5 & 128.4 $\pm$ 7.7 & 47.2 & 45.5 $\pm$ 0.14  \\
    NGC~4526 & rotating disk & 0.509 $\pm$ 0.002 & 113.2 $\pm$ 0.6 &  8.59 $\pm$ 0.01 & 17.5 & 111.3$\pm$ 0.5 & 1.9 & 2.68 $\pm$ 0.08  \\
    NGC~4643 & rotating disk & 0.153 $\pm$ 0.005 & 53.6 $\pm$ 0.5 & 7.27 $\pm$ 0.12 & 7.80 & 48.8 $\pm$ 1.4 & 4.7 & -  \\
    NGC~4684 & filaments & 1.26 $\pm$ 0.01 & 163.8 $\pm$ 0.5 &  7.21 $\pm$ 0.11 & 0.97 & 21.7 $\pm$ 0.9 & 142 & 0.38 $\pm$ 0.06  \\
    NGC~4710 & rotating disk & 0.565 $\pm$ 0.005 & 28.3 $\pm$ 1.5 & 8.72 $\pm$ 0.01 & 5.77 & 26.2 $\pm$ 2.3 & 2.1 & 0.90 $\pm$ 0.06  \\
    NGC~5507 & rotating disk & 0.109 $\pm$ 0.002 & 139.7 $\pm$ 0.9 & $<7.70$ & 5.36 & 59.7 $\pm$ 0.9 & 80 & 0.61 $\pm$ 0.10   \\
    \hline
  \end{tabular}

\end{table*}

\section{Discussion} 
\label{sec:discuss}

Our subset of non-central ETGs from the ATLAS$^{3D}$ catalogue contain warm ionized gas reservoirs that are diverse both in size and morphology. We also find a broad range in 0.5-2.0 keV X-ray luminosities. The overarching question is how can these observed properties illuminate the physical drivers of galaxy evolution in the low-end of the ETG mass-hierarchy. How do these mechanisms compare to those observed in the most massive systems? In this section, we discuss additional analyses of our archival MUSE and \emph{Chandra} data in attempt to identify (1) the connection between the cold molecular gas and warm ionized gas and (2) the origin of the observed cooler gas phases such as cooling of the hot ISM, stellar mass loss, and external accumulation.

\subsection{Cold and warm ionized gas content} 
\label{sec:coldwarm}

\begin{figure*}
    \centering
    \includegraphics[width=\linewidth]{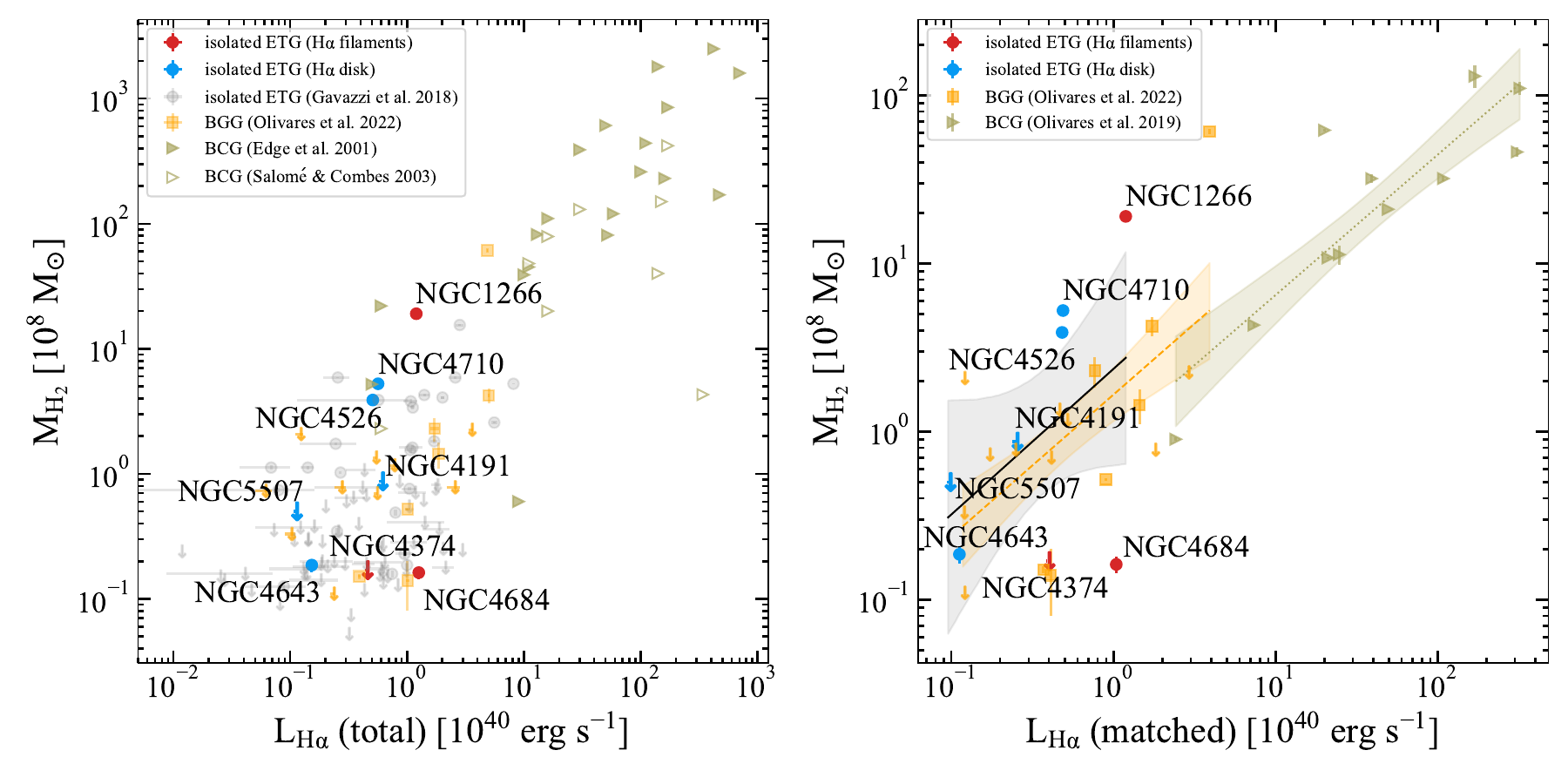}
    \caption{\textbf{Left}: Cold molecular gas mass (vertical axis) plotted against the \emph{total} H$\alpha$ luminosity (horizontal axis). Included are our measurements for our sources with H$\alpha$ filaments (red points) and rotating H$\alpha$ disks (blue points). Upper-limits on the molecular gas masses are depicted by downwards arrows, as with the other over-plotted samples. Gray circles represent the H$\alpha$ luminosity and H$_2$ mass measurements for non-central ETGs reported in \citet{gavazzi2018halpha}. Gold squares represent the H$\alpha$ luminosity measurements for BGGs reported in \citet{olivares2022gas}. Olive-colored triangles represent the measurements for BCGs. Filled and open data points are gathered from \citet{edge2001detection} and \citet{salome2003cold}, respectively. 
    \textbf{Right}: Cold molecular gas mass (vertical axis) plotted against the H$\alpha$ luminosity measured within the \emph{same} aperture size (horizontal axis). Measurements for our sources with H$\alpha$ filaments and rotating H$\alpha$ disks are represented by red and blue points, respectively. Downwards arrows depict upper-limits on the H$_2$ detections (as with the other over-plotted samples). Gold squares represent the H$\alpha$ luminosity measurements for BGGs reported in \citet{olivares2022gas}. Olive-colored triangles represent measurements of BCGs reported in \citet{olivares2019ubiquitous}. Trend lines along with uncertainties are included for each of the three samples. The solid black line represents our non-central ETG sample and is described by the relation:  $\textrm{log(M$_{\mathrm{H_2}}$)} = (0.87\pm1.20)\textrm{log($L_{\mathrm{H\alpha}}$)} + (0.37\pm0.69)$. The dashed gold line corresponds to BGGs and is described by the relation:  $\textrm{log(M$_{\mathrm{H_2}}$)} = (0.84\pm0.30)\textrm{log($L_{\mathrm{H\alpha}}$)} + (0.22\pm0.16)$. The dotted olive-colored line corresponds to BCGs and is described by the relation:  $\textrm{log(M$_{\mathrm{H_2}}$)} = (0.83\pm0.21)\textrm{log($L_{\mathrm{H\alpha}}$)} + (0.02\pm0.37)$.}
    \label{fig:LHaMH2}
\end{figure*}

Masses of the cold gas reservoirs within our sample of non-central ETGs are provided in Table~\ref{tab:data}. This cold molecular gas is traced by the CO emission line, which was observed with the IRAM 30m telescope as described in \citet{young2011atlas3d}. A wide range of cold molecular gas masses are found in these systems ($\sim$ 0.2$-$20$\times10^9$ M$_\odot$). Fig.~\ref{fig:LHaMH2} (left panel) compares these gas masses with the total H$\alpha$ luminosity. Our sample is grouped based on their either filamentary (red data points) or disky (blue data points) H$\alpha$ morphology.

As a comparison, we include H$\alpha$ luminosity measurements from \citet{gavazzi2018halpha} (gray circles), whose sample of non-central ETGs were also comprised of ATLAS$^{3D}$ targets. The H$\alpha$ measurements from \citet{gavazzi2018halpha} were made with the 2.1 m telescope at the San Pedro Martir Observatory, belonging to the Mexican Observatorio Astronómico Nacional. However, this telescope could not independently resolve the H$\alpha$ emission line from the [NII] emission lines. Since their flux estimates pertain to the H$\alpha$+[NII] line, the comparison with our ETGs observed with MUSE is not purely valid. We attempt to ``de-blend" the [NII] emission from the  H$\alpha$+[NII] emission in their measurements by assuming an H$\alpha$ to [NII] ratio of $0.63 \pm 0.05$. This ratio was determined by calculating the H$\alpha$ to [NII] ratio (within the MUSE data) of each of our sample's sources, and then taking the average. Our average ratio is comparable to that found in \citet{olivares2022gas}. 

We compare this relation with higher mass systems such as group-centered and cluster-centered galaxies. BGGs are included as gold squares, with H$\alpha$ luminosity measurements from \citet{olivares2022gas}. Molecular gas masses of the BGG sample (in both panels) are from either IRAM 30m or APEX
telescope observations, as reported in \citet{o2015cold, o2018cold}. BCGs are over-plotted as olive-colored triangles. Both the H$\alpha$ luminosity and molecular gas masses are from \citet{edge2001detection} or \citet{salome2003cold}.

The right panel of Fig.~\ref{fig:LHaMH2} similarly plots the parameters from the left panel. However, here, the luminosity of the H$\alpha$ nebulae is measured within the same size region used to estimate the CO emission. The region we consider for our sample of non-central ETGs is 22$^{\prime\prime}$ around the center of the galaxy \citep{young2011atlas3d}. We similarly show these matched extraction size measurements for BGGs (gold squares) and BCGs (olive-colored triangles). Measurements for the BGGs are from the same references as the left panel. The BCG measurements here are reported in \citet{olivares2019ubiquitous}, where the H$\alpha$ luminosity and molecular gas masses were determined using MUSE and ALMA observations, respectively. The trend appears to tighten when evaluating the molecular gas mass and H$\alpha$ luminosity within the same region for both BGGs and BCGs. However, this change in the relationship is subtle in our sample of non-central ETGs (non-central ETGs from \citet{gavazzi2018halpha} are omitted from this right panel, as the matched-aperture H$\alpha$ measurements were unavailable).

We used the \verb|LINMIX| \citep{kelly2007some} Bayesian linear fitting routine to quantify the matched aperture $L_{\mathrm{H\alpha}}$$-$M$_{\mathrm{H_2}}$ relation. In Fig.~\ref{fig:LHaMH2} (right) the trend lines along with uncertainties are included for each of the three samples. On average, the H$\alpha$ luminosity of non-central ETGs, BGGs, and BCGs appears to follow a clear positive correlation with their molecular gas mass. While their slopes are similar, non-central ETGs have a significantly larger scatter compared to the BGG and BCG sample. This hints that the viable formation channels for the warm ionized gas within non-central ETGs are more diverse. 

\citet{olivares2019ubiquitous} reported that in BCGs, the cold gas was co-spatial and co-moving with the warm ionized gas, giving further evidence that these two gas phases are connected (also see \citet{tremblay2018galaxy}). A similarly positive correlation represented by the non-central ETGs studied here and in \citet{gavazzi2018halpha}, indicates that this linkage between phases exists even in the lower mass ETGs. Detailed spatial and kinematic analysis of the cold gas is required to determine whether the two phases are co-spatial and co-moving like they are in many BGGs and BCGs.

\begin{figure*}
    \centering
    \includegraphics[width=\linewidth]{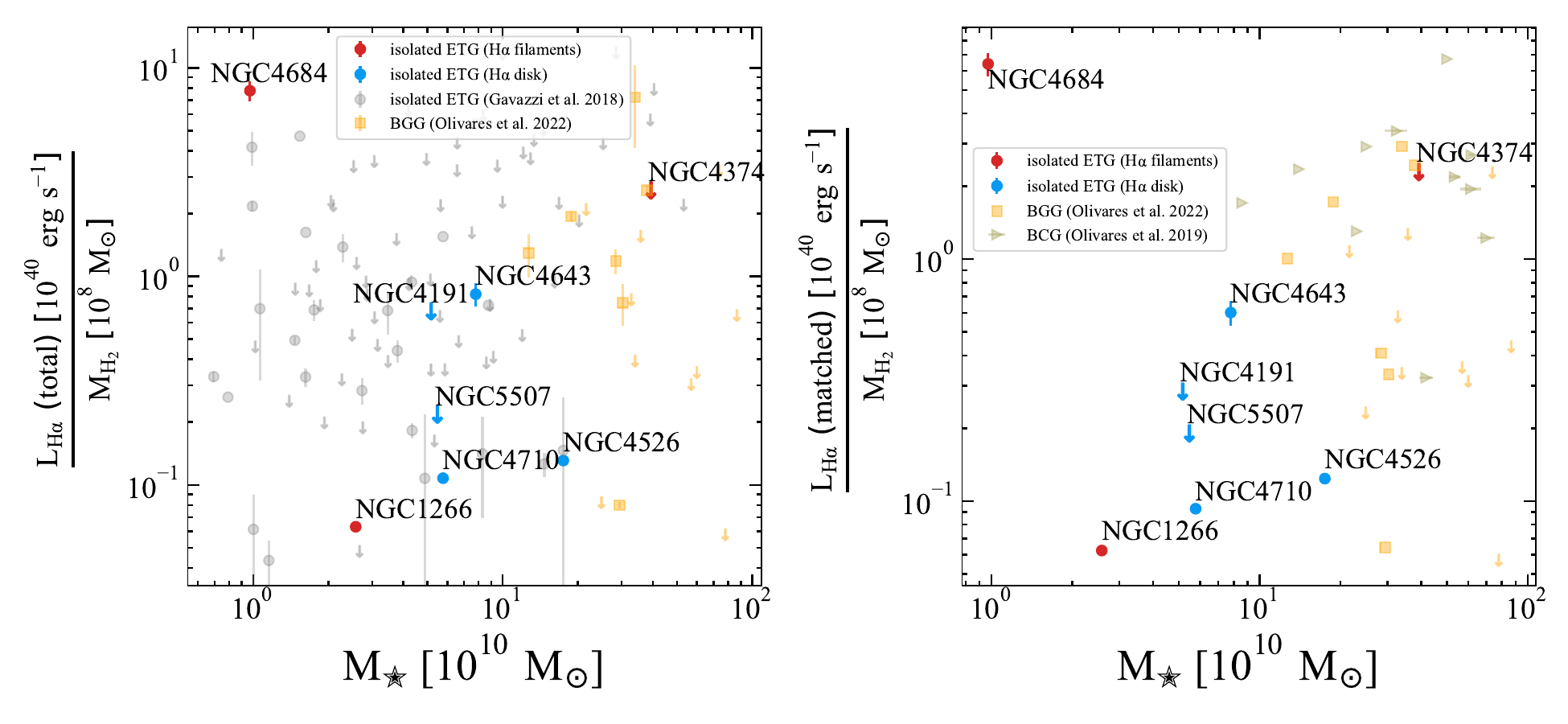}
    \caption{\textbf{Left}: The ratio between the total H$\alpha$ luminosity and cold molecular gas mass is plotted on the vertical axis. Included are our measurements for sources with H$\alpha$ filaments (red points) and rotating H$\alpha$ disks (blue points). We over plot the corresponding values for additional non-central ETGs (gray circles) as well as group-centered galaxies (gold squares). \textbf{Right}: The ratio between the H$\alpha$ luminosity and cold molecular gas mass within the same size region is now plotted on the vertical axis. We again over plot these values for BGGs (gold squares). Measurements for BCGs are also included as olive-colored triangles. See Section~\ref{sec:coldwarm} for the full description of all samples included.}
    \label{fig:ratio}
\end{figure*}

Fig.~\ref{fig:ratio} assesses how the ratio between the H$\alpha$ luminosity varies with stellar mass for these samples, similarly distinguishing between H$\alpha$ luminosities extracted from the entire galaxy (left panel) vs. the CO emission-matched aperture (right panel). Stellar masses for our sample, as well as the sample described in \citet{gavazzi2018halpha}, were gathered from \citet{cappellari2013atlas3d} who estimated the mass-to-light ratio of the best-fitting Jeans Anisotropic Model (JAM; within a region of radius 1.0 R$_e$), as well as the analytic luminosity of the Multi-Gaussian Expansion (MGE) model in the SDSS r-band within the same region. The product of these calculations is the total mass contained within a 1.0 R$_e$ region, which is predictably dominated by the stellar population. H$\alpha$ luminosities of the BGGs (gold squares) in both panels are from \citet{olivares2022gas}. Stellar mass estimates for the BGGs in both panels are from \citet{kolokythas2022complete}. Their stellar mass estimates were derived using a conversion between K-band magnitude and stellar mass from ATLAS$^{3D}$ galaxies. BCGs (olive-colored triangles) are only included in the right panel. Original stellar mass estimates for the BCGs are from a variety of works as described in \citet{olivares2019ubiquitous}. Non-central ETGs from \citet{gavazzi2018halpha} are omitted in the right panel since the matched aperture size H$\alpha$ measurements are unavailable. 

There is no immediate relation found in Fig.~\ref{fig:ratio}. Our sample has typical gas content and stellar mass when compared to other non-central ETGs from \citet{gavazzi2018halpha}, indicating that while the sample size is small, our sources are representative of other nearby non-central ETGs. Comparing the H$\alpha$ and H$_2$ properties with the dark matter halo mass may reveal a tighter correlation. 



\subsection{Formation channels of cool gas phases}
\label{sec: origin}

Cold gas occupying the most massive ETGs, such as cluster-centered galaxies, is believed to form due to the cooling of their hot ICM. 
Galaxies that reside in the center of galaxy groups with an X-ray bright IGrM are also believed to have their cold gas form from IGrM cooling, although, this is not necessarily their primary formation channel. Groups dominated by lower mass spiral galaxies, such as our local group, can have an X-ray faint IGrM where cooling is not expected to be the primary source of their H$\alpha$ gas. 
Naturally, this motivates the pursuit of various formation channels in less-shielded systems, such as non-central ETGs that do not sit at the center of clusters nor groups. In the following subsections, we explore the feasibility of these formation channels within our sample of non-central ETGs.

\subsubsection{Cooling of the hot ISM} 
\label{sec:cooling}

In high mass systems, such as group and cluster-centered galaxies, a significant portion of the observed cold gas is believed to have originated from thermally unstable cooling of the hot, X-ray emitting gas. Without some response to balance this cooling, such as AGN feedback, runaway cooling would produce a continuous stream of in-falling gas (i.e. cooling-flows). Runaway cooling is inconsistent with observations (leading to thus the ``cooling-flow problem" \citep{lea1973thermal}). Constraining these cold gaseous properties in ETGs clearly opens the door to probing the AGN feedback loop, since this cold gas is believed to be the fuel for the central AGN.

\begin{figure}
	
	\includegraphics[width=\columnwidth]{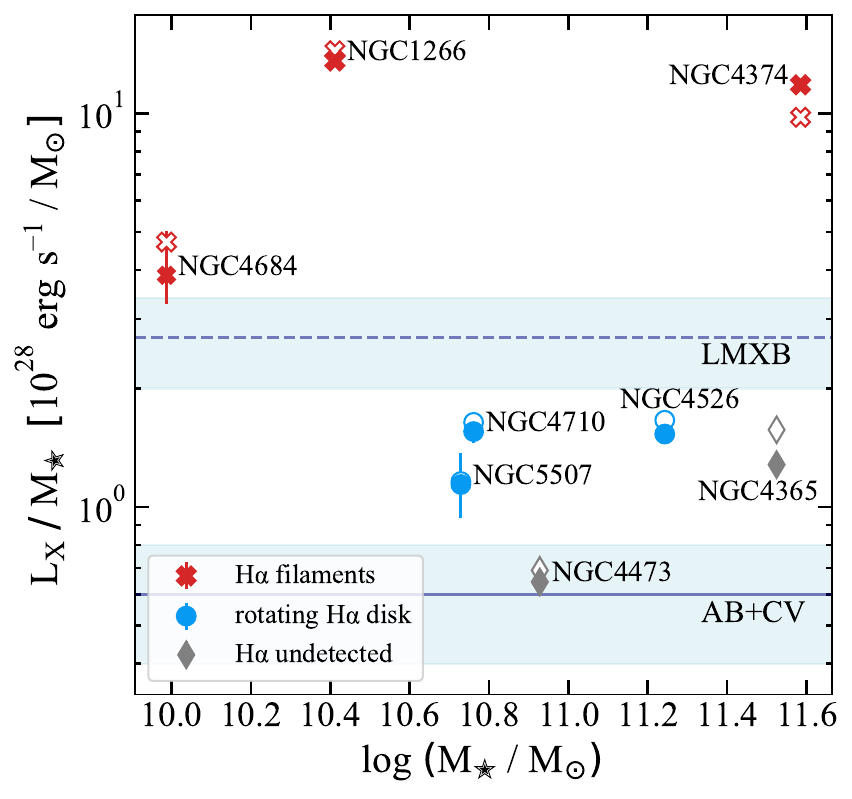}
    \caption{0.5-2.0 keV luminosity, divided by the stellar mass (vertical axis) versus the stellar mass (horizontal axis). Non-central ETGs with filamentary H$\alpha$ gas are depicted by red x's. Whereas sources with rotating H$\alpha$ disks are shown with blue circles. Non-central ETGs in our sample that have X-ray observations, but undetected in H$\alpha$, are depicted as gray diamonds. Filled in data points correspond to $L_X$ estimates assuming a Plasma/APEC model (see Sec.~\ref{subsec:chandra} for a discussion on the choice of $kT$ and $Z$). The open data points correspond to $L_X$ estimates assuming a power-law model ($\Gamma=1.7$). The dashed and solid blue lines indicate the estimated 0.5-2.0 keV emission from unresolved LMXBs and ABs+CVs, respectively. The value and uncertainty of these two estimates are given in \citet{hou2021x}. All non-central ETGs in our sample with H$\alpha$ filaments and X-ray data reveal an X-ray luminosity that exceeds both estimates from the stellar population. While the other sources, hosting rotating H$\alpha$ disks, do not (including the H$\alpha$ non-detections). We expect the former ETGs to host truly diffuse, hot gas.}
    \label{fig:Lx}
\end{figure}

Capturing the diffuse hot gas 
in non-central ETGs, is difficult. The X-ray emission in these systems is typically dominated by discrete sources such as active binaries (ABs), cataclysmic variables (CVs), and low-mass X-ray binaries (LMXB). 
The existence of hot gas occupying the ISM of non-central ETGs can be probed by assessing how the expected X-ray emission from discrete sources compares to the observed total X-ray luminosity. \citet{hou2021x} provides estimates of the average 0.5-2.0 keV X-ray emissivity of ABs, CVs, and unresolved LMXBs in low-intermediate mass ETGs. Fig.~\ref{fig:Lx} shows how the X-ray luminosity, normalized by the stellar mass (for sources with available X-ray data), compare with the approximate emission governed by those relations. 
The total 0.5-2.0 keV luminosity exceeds the X-ray luminosity estimates for both the AB+CVs and unresolved LMXBs within all non-central ETGs that display filamentary H$\alpha$ morphologies. This excess in X-ray emission indicates the presence of an additional X-ray component in these sources which we posit to be
the hot gaseous phase. Though, caution must be taken with the classification of NGC~4684 being hot gas rich due to the poor quality of existing data. Its hot gas content remains to be verified with future on-axis observation (see Sec.~\ref{subsec:chandra} for further details). The X-ray luminosity of the other ETGs in our sample that have rotating H$\alpha$ disks (NGC~4526, NGC~4710, and NGC~5507) does not entirely surpass the X-ray luminosity expected from the discrete sources. Nor do the two sources in our sample (NGC~4365 and NGC~4473) that went undetected in H$\alpha$.

Such an interpreted connection between the H$\alpha$ morphology and presence (or lack) of diffuse hot gas tells a compelling story regarding the possible origin of these H$\alpha$ filaments. In cool-core clusters with low core entropy and short central cooling times, rich with diffuse hot gas, filamentary H$\alpha$ networks are observed (\citet{voit2015cooling}, \citet{o2018cold}), and are believed to be the product of cooling of the hot ICM. Given that similar filamentary structures are observed only in our systems with diffuse hot gas, it is natural to infer that cooling of the hot ISM is a significant source of their cool phases.

If these filaments are truly formed from cooling induced by thermal instabilities, then prominent AGN activity is anticipated in NGC~1266, NGC~4374, and NGC~4684. Early radio observations of NGC~4374 reported in \citet{jenkins1977observations} display bipolar jets extending from its core, and it is classified as a Fanaroff Riley I (FRI) radio galaxy based on the classification scheme in \citet{fanaroff1974morphology}. \citet{alatalo2011discovery} found a molecular outflow in NGC 1266 which they deduced to be powered by an AGN (see Section~\ref{sec:outflows} for further discussion). Such conclusions agree with \citet{nyland2016atlas3d} who found indicators of AGN activity within these two sources, as well as NGC~4684. This study found that these three ETGs displayed radio morphologies with resolved, extended emission indicative of an AGN jet.

Further analysis of thermodynamic properties such as entropy, cooling time, etc., are required to assess the feasibility of cooling from the hot halo. This is well-motivated as filaments are observed in cool-core clusters with central regions displaying low entropy and short cooling times (see \citet{hogan2017onset,pulido2018origin} and references therein). However, calculation of these properties require deeper X-ray coverage than what is available for most of our systems.


\subsubsection{Stellar mass loss} 
\label{sec:mdot}

Stellar mass loss is another candidate for the origin of the cold gas. In this scenario, the majority of the stellar ejecta is expected to be thermalized by gas shocks \citep{parriott2008mass, bregman2009mass,voit2011fate}. However, a considerable amount of the ejecta will not be thermalized and will settle onto the central disk.

If true, one would expect kinematic alignment and overlap between the rotating H$\alpha$ disk and the rotating stellar disk due to conservation of angular momentum. This kinematic alignment can be quantified by comparing the position angle (PA; \citet{krajnovic2006kinemetry}) of the two rotating populations. The PA describes the orientation of the bulk motion of the H$\alpha$ gas and stellar population. This provides a consistent geometric interpretation of the motion, specifically as the angle between the north and the position of where the maximum, positive velocity occurs.

\begin{figure}
	\includegraphics[width=\columnwidth]{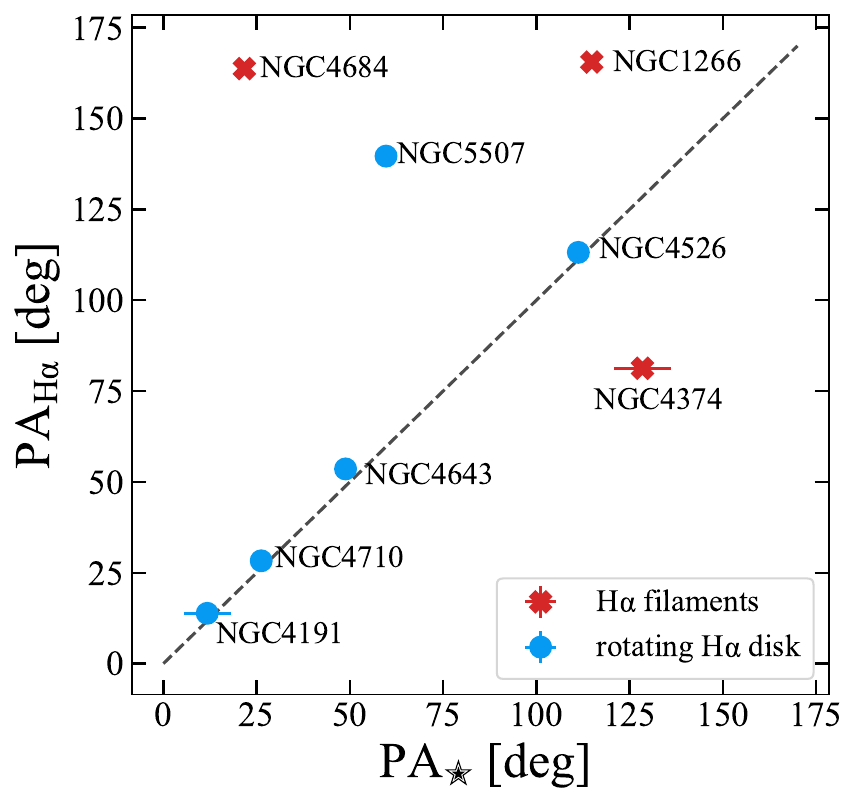}
    \caption{The H$\alpha$ kinematic PA (vertical axis) vs. the stellar kinematic PA. Non-central ETGs with filamentary H$\alpha$ gas are depicted by red x's. Sources with rotating H$\alpha$ disks are shown with blue circles. The dashed gray line represents the one-to-one line. All filamentary H$\alpha$ ETGs have decoupled motion between the warm ionized gas and stars ($\Delta$PA $> 30^{\circ}$). Whereas all but one (NGC~5507) sources with rotating H$\alpha$ disks show coupled motion ($\Delta$PA $< 30^{\circ}$) between the warm ionized gas and stars.}
    \label{fig:PA}
\end{figure}

Fig.~\ref{fig:PA} compares the position angle of the H$\alpha$ gas (PA$_{\mathrm{H}\alpha}$) and the stellar disks (PA$_{\star}$) in our sample. The grey dotted-line indicates the one-to-one line. We observe that almost all of our rotating H$\alpha$ disks (with the exception of NGC~5507) have $\Delta \textrm{PA} = | \textrm{PA}_{\mathrm{H}\alpha} - \textrm{PA}_{\star} | < 30^{\circ}$. Similar to \citet{lagos2015origin}, we adopt this value to determine if the stars and gas are kinematically coupled ($\Delta \textrm{PA} < 30^{\circ} $) or decoupled ($\Delta \textrm{PA} > 30^{\circ} $). All three ETGs with H$\alpha$ filaments have a PA$_{\mathrm{H}\alpha}$ that is widely offset from PA$_{\star}$. 

The kinematic position angles between the gas and stars within the ETGs with rotating H$\alpha$ disks (excluding NGC~5507) are comparable, while the filamentary H$\alpha$ gas in ETGs is completely decoupled from the stellar component. The decoupled motion seen in  the filamentary H$\alpha$ sources compliments our discussion in Section~\ref{sec:cooling}, which explains how cooling of the hot ISM could be a possible source of the cold gas in those ETGs. The rotating H$\alpha$ disks that appear coupled to the stars support the stellar mass loss scenario to be a viable origin of the observed cold gas. Although, it is worth mentioning that while NGC~4191 and NGC~4643 appear to have kinematic alignment, it is easy to see in Fig.~\ref{fig:Ha_Stell} that their stars and gas do not entirely overlap. The contribution of stellar mass loss may be less significant in these systems, compared to NGC~4526 and NGC~4710 which have kinematic alignment and fully overlapping stars and gas.

Tracing the origin of cold gas specifically back to stellar mass loss is difficult in ETGs with ordered stellar rotation and hot ISM rotation. In these systems, rotating X-ray halos may condense into colder gas that can lead to the formation of a rotating cold gas disk. The effects of angular momentum conservation can result in a decrease in density of the hot ISM, consequently lowering the host galaxy's X-ray temperature and X-ray surface brightness as demonstrated in numerical simulations by \citet{negri2014effectsa,negri2014effectsb}. This possible decrease in X-ray surface brightness in these systems could make it difficult to detect these rotating hot halos. Such an effect would complicate the precise mapping of the cold gas disks back to stellar mass loss vs. condensation of the rotating hot halo.

\subsubsection{External origin} 
\label{sec:mergers}

In the previous subsections, we discuss the feasibility of cooler gas phases originating from processes that occur within non-central ETGs such as cooling of the hot ISM and stellar mass loss. It is also possible for this multiphase gas to be externally supplied through accretion from the intergalactic medium (IGM), interaction with surrounding satellites, and accumulation via merging with another gas-rich galaxy (wet-merger). Here we discuss the analysis of our MUSE data, as well as past studies, to determine whether external processes could be responsible for the multiphase gas observed in our sample.

In the context of a wet-merger, cold gas can be exchanged and accumulated. Shock-heating would likely cause most of the exchanged gas to add to the hot ISM \citep{cox2006x}. Yet, some of the cold gas may dodge this heating and settle onto the resulting galaxy's rotating gaseous disk. As this gas settles onto the disk, the angular momentum of the disk can be preserved. Whereas if the merger involves a gas-poor galaxy (dry-merger), angular momentum of the disk can be decreased, especially if the galaxy has been involved in multiple dry mergers (\citealp[see][for review]{cappellari2016structure}). 

Previous studies have found that galaxies displaying ordered rotation in their velocity fields typically have large stellar angular momentum (and conversely galaxies with minimal rotation show lower angular momentum) \citep{krajnovic2011atlas3d, loubser2022merger}. This relationship offers insight on the evolutionary channels (i.e. having experienced wet versus dry-mergers) of galaxies that display ordered stellar rotation in their velocity fields, although, a visual classification of the velocity fields is not sufficient to strictly deduce high angular momentum. 

\citet{emsellem2011atlas3d} provides an additional probe as to whether a galaxy is a ``slow" or ``fast" rotator using a relationship between its ellipticity ($\epsilon$) and projected angular momentum ($\lambda$). Here, $\lambda$ is a luminosity-weighted  estimate of the specific angular momentum of the stellar population. It is derived from the first and second stellar velocity moments. Their study found that $\lambda > 0.31\sqrt{\epsilon}$ was a sufficient descriptor of fast rotators. Both parameters for our non-central ETGs are provided in \citet{krajnovic2011atlas3d} and \citet{emsellem2011atlas3d}.

\begin{figure}
	\includegraphics[width=\columnwidth]{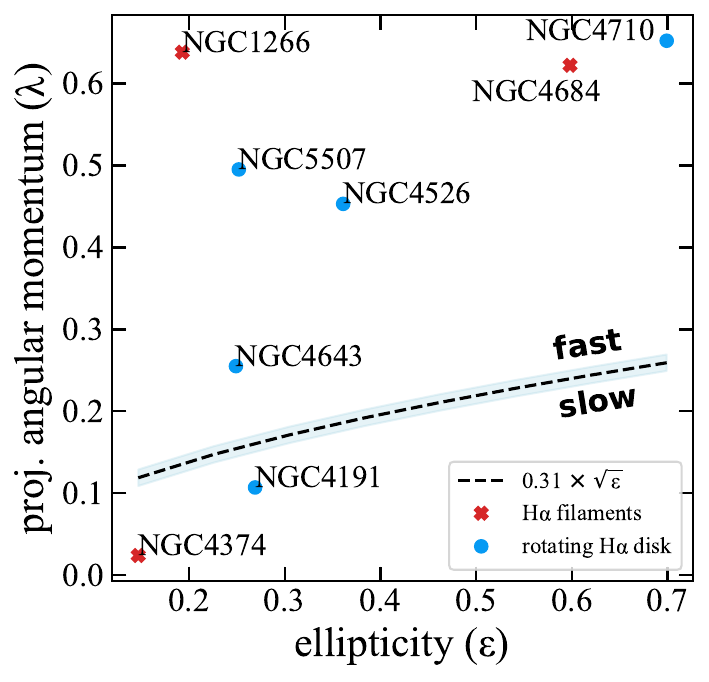}
    \caption{The projected angular momentum (vertical axis) as a function of the global ellipticity (horizontal axis). Red x's indicate our non-central ETGs that have filamentary H$\alpha$ nebulae. Blue circles depict galaxies that host rotating H$\alpha$ disks. The black dashed line is the relation that separates the fast rotators (above dashed line) from the slow rotators (below dashed line).}
    \label{fig:rotation}
\end{figure}


Fig.~\ref{fig:rotation} shows how our sample of non-central ETGs compare to this relation. NGC~4191 and NGC~4374 are the only sources considered to be slow rotators based on this classification scheme. Their stellar velocity maps in Fig.~\ref{fig:Ha_Stell} also do not display clear, ordered rotation.

We find a collection of fast rotators that are divided into two classes. Those with gas and stars that are either kinematically aligned (NGC~4526, NGC~4643, and NGC~4710) or misaligned (NGC~1266, NGC~4684, and NGC~5507). For both of these groups, the fast rotation indicates that it is plausible for their cooler gas phases to have partially originated from external accretion events. Although, the distinction between their gas and stars being kinematically aligned or misaligned provides valuable insight on the relevant timescales and conditions possibly required to produce their observed kinematics. Simulations found that misalignment between an ETG’s gas and stars following a major merger can persist for $\sim$ 2 Gyr \citep{van2015creation}. After the merger, continual accretion of gas would dominate the change of the newly formed gas disk’s angular momentum vector. Over time, the accretion rate of the gas would decrease sufficiently for torque induced by stars to begin dominating the angular momentum evolution. At this stage, the angular momentum vectors of the gas would quickly align to those of the stars, thus allowing for co-rotation (or counter-rotation). Although, further accretion of material from the cosmic web and merger tidal remnants, as well as rotation of the hot halo, may increase the relaxation time \citep{davis2016depletion}. This elongated relaxation time could explain why some local fast rotators such as NGC~1266, NGC~4684, and NGC~5507 retain their observed misalignment.

NGC~4191 is a unique source that possibly displays a separate indicator of previous external accretion. Surrounding the optical center is a ring consisting of star-forming, H$\alpha$ gas clumps. Such H$\alpha$ rings have been similarly observed in BGGs \citep{loubser2022merger, olivares2022gas}. These rings are believed to be products of external processes such as wet mergers or tidal interactions \citep{mapelli2015building}, which would consequently bring in new gas. We would thus expect NGC~4191 to sit above the $\lambda-\epsilon$ cutoff (suggesting that the source possibly experienced recent external accretion), contrary to what we find Fig.~\ref{fig:rotation}. However, it is worth mentioning that the projected angular momentum values were generated using the SAURON instrument with a small field-of-view (33$^{\prime\prime}$$\times$41$^{\prime\prime}$), and thus projection affects may cause an underestimation of the projected angular momentum. Thus we cannot rule out the possibility of external accretion to be a formation channel of the multiphase gas in NGC~4191. Stellar mass loss would not produce the star-forming rings, leading to the assumption that external accretion is instead the primary formation channel.

The rotating H$\alpha$ disks in NGC~4643 and NGC~5507, as observed in Fig.~\ref{fig:Ha_Stell}, display extended structures which may be further signs of past interaction, such as tidal stripping of a nearby satellite. Although their gas and stars show comparable position angles, their distributions do not entirely overlap, indicating that external accretion is likely the primary origin in these two systems.

\subsubsection{NGC~1266 and its multiphase outflows} 
\label{sec:outflows}

NGC~1266 is a nearby ($D=29.9$ Mpc) S0 galaxy that displays unique multiwavelength properties. \citet{alatalo2011discovery} found an unresolved radio point source along with an extended jet within its 1.4 and 5 GHz continuum image. They found X-ray emission dominated by thermal bremsstrahlung as well as hard X-ray photons co-spatial with the radio source. Such signatures reveal NGC~1266 to be a host of an obscured AGN.

This nearby AGN holds a distinct rank amongst the ATLAS$^{3D}$ sample. \citet{young2011atlas3d} reported a highly compact molecular gas distribution, with an average gas surface density ($\sim 230$ M$_{\odot}$ pc$^{-2}$) and mass ($\sim 19 \times 10^8 $ M$_{\odot}$) greater than all other ETGs in their sample. The compact molecular gas was similarly detected in \citet{alatalo2011discovery} in the form of a rotating disk (60 pc radius) engulfed within a small, diffuse reservoir. Surprisingly, an additional molecular gas component was found out-flowing normal to the compact rotating disk.

SFR estimates in NGC~1266 were found to be insufficient for stellar outflows to have ejected this molecular gas leaving outflows from its verified AGN to be the lead candidate.  Depicted in Fig.~\ref{fig:Ha_Stell} (bottom row, left panel), our MUSE analysis reveals H$\alpha$ gas occupying the inner-kpc region, with a bright, irregular whirl south of the SMBH. Furthermore, we find high H$\alpha$ velocities and velocity dispersions within the nuclear region of NGC~1266. These features are indicative of outflows, and are corroborated by \citet{davis2012gemini}. There, the authors carried out a study to probe this outflow mechanism, where they used GMOS and SAURON IFU data to interpret the ionized gas content within NGC~1266. 

The position of this ionized gas feature matches with the AGN radio jet as well as the outflow of the molecular gas. \citet{davis2012gemini} showed the outflow velocity of both the neutral and ionized gas lead to similar times for when the outflow could have occurred, giving further evidence that both components were ejected by the same mechanism. 

The unique nature of NGC~1266 complicates, as well as motivates, the search for the origin of its gaseous species. \citet{alatalo2011discovery} notes the puzzling existence of its compact molecular gas core. \citet{alatalo2014suppression} proposed that NGC~1266 experienced a minor merger 0.5 Gyr ago. During this merger, a burst of SF was triggered, yet a significant portion of molecular gas was left over as it failed to condense and form stars. This supports our results presented in Fig.~\ref{fig:rotation}, indicating that NGC~1266 likely encountered a previous wet merger, as it has a noticeably high angular momentum (see discussion in Section~\ref{sec:mergers}). In this process, cold gas could be accumulated which provides a reasonable origin for the compact molecular gas disk.

While a past merger was likely the source for some of its molecular gas, the multiphase outflow driven by the AGN may explain the other gas phases observed in NGC~1266. \citet{davis2012gemini} showed that SF cannot be the dominant excitation mechanism of the optical lines. Rather, the emission from ionized gas in NGC~1266 is predominately caused by fast shocks from the radio jet permeating through the ISM. This places the origin of NGC~1266’s warm gas in a class of its own. Such an association between the outflow and the ionized gas is further supported by the high shock velocities being comparable to those found in the outflow.
The ISM can also be heated by gas shocks, causing the gas to shine bright in the X-ray band. The sharp edges within NGC~1266's X-ray image are likely tracing this feature of shock-heating.

\section{Conclusions} 
\label{sec:conclusion}

The multiphase gaseous properties of a small, yet diverse sample of non-central ETGs are presented in this paper. Both H$\alpha$ filaments and rotating disks revealed by MUSE are found within 8/15 non-central ETGs in our sample. This is similar to the H$\alpha$ morphologies found within BGGs, although rotating H$\alpha$ disks seem to be slightly more frequent in non-central ETGs. There is a clear dichotomy between the morphologies of non-central ETGs and BCGs, as the H$\alpha$ reservoirs in BCGs are mainly filamentary. 
Using non-central ETGs as test beds for interpreting the origins of multiphase gas offers a comparison to higher mass systems, which have been intensively studied. We find that there are various formation channels for the observed cool gas, each providing insight on the dynamical state of these systems. The above analysis allows us to conclude:

\begin{enumerate}

    \item NGC~1266, NGC~4374, and NGC~4684 host H$\alpha$ filaments. NGC~1266's H$\alpha$ emission is compact with knotty clumps, while the H$\alpha$ gas in NGC~4374 and NGC~4684 forms more extended, branch-like structures. Each of these sources show signs of containing hot, diffuse gas. It is likely that cooling of their hot ISM is a dominant origin of their observed cooler gas phases. We emphasize that while our analysis with the existing \emph{Chandra} observation supports NGC~4684 being hot gas rich, future on-axis observations are needed to confirm its hot ISM properties and the origin of its multiphase gas.

    \item NGC~4191, NGC~4526, NGC~4643, NGC~4710, and NGC~5507 all contain rotating H$\alpha$ disks. Those with available X-ray data (NGC~4526, NGC~4710, and NGC~5507) do not show signs of hosting significant hot gas reservoirs. 
    
    \item NGC~4526 and NGC~4710 show ordered stellar rotation that is kinematically aligned ($\Delta$PA $< 30^{\circ}$) and overlaps with their warm ionized gas disks. Both ETGs are also classified as fast stellar rotators. It is expected that a mixture of stellar mass loss and external accretion is the dominant avenue for their cool gas.

    \item NGC~4191 has a clumpy, potentially star-forming H$\alpha$ ring. Stellar mass loss is not expected to produce the observed H$\alpha$ ring. NGC~4643 and NGC~5507 are fast stellar rotators, and their rotating H$\alpha$ disks display extended structures along their outskirts. The gas and stars within NGC~5507 have widely offset kinematic PAs, and even though the two populations within NGC~4643 have similar PAs, their distributions do not entirely overlap. It is likely that the cool gas in NGC~4191, NGC~4643, and NGC~5507 primarily formed through external processes such as wet mergers, tidal interactions, or IGM accretion as opposed to cooling or stellar mass loss.

    
    \item While cooling of the hot gas and external accretion likely formed the cooler phases observed in NGC~1266, this non-central ETG is distinct in that its AGN jet drove a multiphase outflow. This outflow shows evidence of the surrounding ISM being shock heated, which could have added to both the warm ionized gas and hot gas constituting its ISM.
    

\end{enumerate}

\section*{Acknowledgements}
The authors thank the anonymous reviewer for their helpful comments on the manuscript.
R.E., V.O., and Y.S. acknowledge support by NSF grant 2107711, NRAO grant SOSPA9-006, Chandra X-ray Observatory grants GO1-22126X, GO2-23120X, G01-22104X, NASA grants 80NSSC21K0714 and 80NSSC22K0856. Y.L. acknowledges financial support from NSF grants AST-2107735 and AST-2219686, NASA grant 80NSSC22K0668, and Chandra X-ray Observatory grant TM3-24005X. 

\section*{Data Availability}

Data described within this article will be shared on request to the corresponding author.



\bibliographystyle{mnras}
\bibliography{MAIN} 








\bsp	
\label{lastpage}
\end{document}